\documentclass[12pt]{article} %,a4paper
\usepackage{graphicx}
\usepackage{amsmath}
\usepackage{amssymb}
\allowdisplaybreaks

 \usepackage{cite}
 \usepackage{hyperref}
 \usepackage{youngtab}
\usepackage[height=8.8in,width=6.45in]{geometry}
\numberwithin{equation}{section}
\def\Nequals#1{$\mathcal{N}{=}#1$}

\def\mm{\mathfrak{m}}

\def\CB{{\cal B}}

\def\CH{{\cal H}}

\def\CM{{\cal M}}
\def\CN{{\cal N}}

\def\beq#1\eeq{\begin{align}#1\end{align}}

\def\ket#1{\left| #1 \right\rangle}
\def\bra#1{\left\langle #1 \right|}
\def\vev#1{\left\langle #1 \right\rangle}

\def\U{\mathrm{U}}
\def\SU{\mathrm{SU}}

\def\SU{\mathrm{SU}}

\def\tr{\mathop{\mathrm{tr}}}
\def\diag{\mathop{\mathrm{diag}}}
\def\rank{\mathop{\mathrm{rank}}}

\begin{document}

\begin{titlepage}

\begin{flushright}
IFT-UAM/CSIC-14-107\\
IPMU-14-0325\\
UT-14-45\\
\end{flushright}

\vskip 2cm

\begin{center}
{\Large \bfseries
Mass-deformed $T_N$ 
as a linear quiver
}

\vskip 1.5cm
  Hirotaka Hayashi$^1$, Yuji Tachikawa$^2$ and Kazuya Yonekura$^3$
\vskip 1.0cm

\begin{tabular}{ll}
$^1$ & Instituto de F\'\i sica Te\'orica UAM/CSIS, \\
&Cantoblanco, 28049  Madrid, Spain \\
$^2$  & Department of Physics, Faculty of Science, \\
& University of Tokyo,  Bunkyo-ku, Tokyo 133-0022, Japan, and\\
  & Institute for the Physics and Mathematics of the Universe, \\
& University of Tokyo,  Kashiwa, Chiba 277-8583, Japan\\
$^3$ & School of Natural Sciences, Institute for Advanced Study,\\
& Princeton, NJ 08540, United States of America
\end{tabular}

\vskip 1cm

\textbf{Abstract}

\end{center}

\medskip
\noindent
The $T_N$ theory is a non-Lagrangian theory with $\SU(N)^3$ flavor symmetry. 
We argue that when mass terms are given so that two of $\SU(N)$'s are both broken to $\SU(N{-}1)\times \U(1)$, it becomes $T_{N{-}1}$ theory coupled to an $\SU(N{-}1)$ vector multiplet together with $N$ fundamentals.  This implies that when two of $\SU(N)$'s are both broken to $\U(1)^{N{-}1}$, the theory becomes a linear quiver. 

We perform various checks of this statement, by using the 5d partition function, the structure of the coupling constants, the Higgs branch, and the Seiberg-Witten curve.  We also study the case with more general punctures. 

\end{titlepage}

\setcounter{tocdepth}{2}
\tableofcontents

%%%%%%%%%%%%%%%%%%%%%%%%%%%%%%%%%%%%%%%%%%%%%

\section{Introduction and Summary}
In the last few years, strongly-coupled superconformal field theories (SCFT) that do not admit any obvious Lagrangian description in the ultraviolet (UV) play more and more important roles in our understanding of the supersymmetric dynamics and dualities.  In 4d, they are sometimes realized as a subcomponent of strongly-coupled limits of Lagrangian theories \cite{Argyres:2007cn,Argyres:2007tq};  in 5d, they are often conjectured to exist as ultraviolet completions of Lagrangian theories \cite{Seiberg:1996bd,Intriligator:1997pq}.  They can often be constructed using superstring theory and M-theory. 

Among these SCFTs, a central role is played by the so-called $T_N$ theory.  The 4d version, originally introduced in \cite{Gaiotto:2009we,Gaiotto:2009gz}, is an \Nequals{2} superconformal theory with $\SU(N)^3$ flavor symmetry, that arises as the four-dimensional limit of the 6d \Nequals{(2,0)} theory of type $\SU(N)$ on a sphere with three full punctures.  The 5d version was soon introduced in \cite{Benini:2009gi}, as a superconformal theory living on the intersection of $N$ D5-branes, $N$ NS5-branes and $N$ (1,1) 5-branes, and its compactification on $S^1$ gives back the 4d version.

Due to its intrinsic importance, the properties of the $T_N$ theory have been studied in earnest. For example, the partition function of the 4d version on $S^1\times S^3$ was found in \cite{Gadde:2011ik,Gadde:2011uv} using the relation to the 2d topological quantum field theory; that of the 5d version on $S^1\times S^4$ was found in \cite{Hayashi:2013qwa,Bao:2013pwa} using the topological vertex formalism; many of the Higgs branch chiral ring relations were worked out in \cite{Gadde:2013fma,Maruyoshi:2013hja}. 
The theory can be deformed by giving vacuum expectation  to the Higgs branch operators so that we have more general SCFTs labeled by three Young diagrams each with $N$ boxes. The 4d versions are sometimes called the tinkertoys and extensively studied starting from \cite{Chacaltana:2010ks}, and some of their chiral ring relations have been analyzed \cite{McGrane:2014pma}. 

In this paper, we study a different type of deformations, namely by mass terms.  As $T_N$ theories have the flavor symmetry $\SU(N)_A\times \SU(N)_B\times \SU(N)_C$, the mass terms take values in three traceless complex-valued $N\times N$ matrices $\mm_{A,B,C}$ , that are hermitian in the case of 5d version.  The effect of the mass terms when they are nilpotent was studied in \cite{Gadde:2013fma,Maruyoshi:2013hja}, and therefore our aim here is the case when they are diagonalizable. 

We will claim that the mass deformation of the two $\SU(N)$ flavor symmetries by diagonal mass matrices makes the $T_N$ theory flow to a linear quiver theory. This fact and its generalization were observed in \cite{Hayashi:2013qwa, Aganagic:2014oia}, but it was unclear whether the gauge groups are unitary gauge groups or special unitary gauge groups. We will argue that the gauge groups are special unitary groups and there are additional hypermultiplets at the end of the quiver compared to \cite{Hayashi:2013qwa, Aganagic:2014oia}.

\paragraph{Basic statement.} Our basic claim, both in 5d and in 4d, is then the following: let us give mass terms to $\SU(N)_B$ and $\SU(N)_C$ such that they are both broken to $\SU(N{-}1)\times \U(1)$. More explicitly, take the mass terms to be \begin{equation}
\mm_A=0,\quad
\mm_B = m_B \diag(1,1,\ldots, 1-N), \quad
\mm_C = m_C \diag(1,1,\ldots, 1-N).
\end{equation} This triggers a renormalization group (RG) flow, and the infrared limit is described by the following theory: \begin{equation}
[\SU(N)_A]-\SU(N{-}1)-T_{N{-}1}.\label{eq:coupled}
\end{equation}
Here,  the $T_{N{-}1}$ theory is coupled to an $\SU(N{-}1)$ gauge multiplet, that is also coupled to a bifundamental of $\SU(N{-}1)\times \SU(N)_A$.
In \eqref{eq:coupled} the brackets are placed around $\SU(N)_A$ to emphasize that it is a flavor symmetry.
In the 5d version, the mass $m_{\rm bif}$ of the bifundamental  and the gauge coupling $8\pi^2/g^2$ of the $\SU(N{-}1)$ are given by
\beq
m_{\rm bif}&=m_{B}+m_{C},
& \frac{8\pi^2}{g^2} &= \frac{N}{2} (m_B -m_C),
\eeq and the Chern-Simons level of the $\SU(N{-}1)$ gauge group is zero. 

For example, take $N=3$. This is the 5d version of the $E_6$ theory of Minahan and Nemeschansky. After the mass deformation, we have $\SU(2)$ coupled to $T_2$ and three flavors. Since $T_2$ is equivalent to two flavors of $\SU(2)$, the infrared theory is just $\SU(2)$ with five flavors. This is the setup originally found by Seiberg \cite{Seiberg:1996bd}, where this class of 5d SCFTs was first discussed.

We can also consider an even simpler case of $N=2$. Recall that the $T_2$ theory is just the tri-fundamental of $\SU(2)_A\times \SU(2)_B\times\SU(2)_C$. 
Giving masses $(m_B,-m_B)$ and $(m_C,-m_C)$ to $\SU(2)_{B,C}$, we have two flavors of $\SU(2)_A$, with masses $m_B+m_C$ and $m_B-m_C$.  In the limit $|m_B+m_C| \ll |m_B-m_C|$, we just have one flavor of $\SU(2)_A$. This is the bifundamental of $\SU(2)\times \SU(1)$, with mass $m_B+m_C$. 

\paragraph{Recursive application.} Recursively applying this procedure, we immediately find that the infrared outcome of a more general  mass deformation given by \begin{equation}
\mm_A=0,\quad
\mm_B =  \diag(m_{B,1},m_{B,2},\ldots, m_{B,N}), \quad
\mm_C = \diag(m_{C,1},m_{C,2},\ldots, m_{C,N})  \label{generalmass}
\end{equation} is a linear quiver theory of the form \begin{equation}
[\SU(N)_A]-\SU(N{-}1)-\SU(N{-}2)-\cdots -\SU(2)-\SU(1)
\end{equation} where  groups enclosed in the brackets are flavor symmetries, other groups are gauged, and we have bifundamental hypermultiplets  for each  consecutive pair of groups. In the 5d version, all the Chern-Simons levels are zero. The same statement recently appeared in \cite{Bergman:2014kza}.
 It turns out that ``$\SU(1)$" should be formally understood as an additional hypermultiplet charged under the $\SU(2)$
in addition to the bifundamental of $\SU(2)-\SU(1)$. This can be seen by stopping the recursive process at $T_2$.

These statements can be easily generalized, by giving nilpotent vevs to the chiral operators in the adjoint of the $\SU(N)_A$ flavor symmetry.
This process is often called the `partial closing of the puncture' in the 4d class S theory,
and we use the same terminology even in the 5d case.

In this language, the $T_N$ theory has three punctures, and in the more general case, 
we start from the theory with two full punctures and a puncture of type $Y=[n_1,n_2,\ldots,n_k]$ with $\sum n_i=N$. We still have the flavor symmetry $\SU(N)_B\times \SU(N)_C$ to which we give masses as in \eqref{generalmass}.  Then we have a quiver theory of the form \begin{equation}
\SU(v_{1})-\SU(v_{2})-\cdots -\SU(v_{N{-}2})-\SU(v_{N{-}1})\label{eq:generalQ}
\end{equation} with additional $w_i$ fundamental hypermultiplets for $\SU(v_i)$, where
$w_k$ is the number of times $k$ appears in the partition $[n_i]$, and $v_i$ are defined by the relation \begin{equation}
v_{N{-}1}=1,~v_N:=0;\quad  2v_i=v_{i-1}+v_{i+1} + w_i\ \text{for $i=2,\ldots,N{-}1$}.\label{eq:generalR}
\end{equation}   
It is interesting to note here that the 3d quiver description of the 3d theory $T_Y(\SU(N))$, introduced originally in \cite{Gaiotto:2008ak}, has the same structure except that the groups are $\U(v_i)$. The reason will be uncovered in Sec.~\ref{sec:3d}.

The quiver \eqref{eq:generalQ} can also be realized as an $A_{N{-}n_1-1}$ class S theory with $N$ simple punctures and one puncture of type
$Y'=[n_2,\ldots,n_k]$.

\paragraph{Organization of the paper.}
In Sections~2, 3 and 4, we perform various tests to check these proposals. The checks given in those sections are mostly independent from each other, and can be read independently, depending on the taste of the reader. 

We start in Sec.~\ref{sec:brane} by  considering the 5d version of the story, where the $T_N$ theory has a construction by a web of branes and the relation to the linear quiver can be most easily seen. 
We recall the method to compute its Nekrasov partition function from the topological vertex, and use it to relate the mass parameters of the $T_N$ to the gauge couplings and the masses of the linear quiver theory.  

Next, in Sec.~\ref{sec:rg}, we perform a field-theoretical analysis to check that under the mass deformation preserving $\SU(N{-}1)_{B,C}$, the $T_N$ theory becomes the coupled theory \eqref{eq:coupled}.  We consider the matching of the operators and of the vacuum moduli spaces, and speculate what happens when $\mm_C=0$.

Then, in Sec.~\ref{sec:qv}, we study the field-theoretical analysis of the relation between the mass-deformed $T_N$ theory and the linear quiver.  In particular, we study the Seiberg-Witten curves and the Higgs branches.  We also analyze the system when  we replace the full puncture carrying $\SU(N)_A$ with a more general puncture. We also perform in Sec.~\ref{sec:3d} the analysis in the 3d version of the $T_N$ theory.

In Appendix~\ref{sec:higgsrelations} we summarize the Higgs branch operators of the $T_N$ theory and their chiral ring relations, some of which are new.

\paragraph{Note added:} Recently there appeared a paper \cite{Bergman:2014kza} where the relation of the mass-deformed $T_N$ theory and the linear quiver of $\SU$ groups was also proposed, and their Sec.~2 and our Sec.~2 have a rather large overlap.  
Also, the relation to the linear quiver of $\U$ groups was already mentioned in \cite{Hayashi:2013qwa} and further studied in detail in \cite{Aganagic:2014oia}.
As our paper appears on the arXiv about two weeks later than \cite{Bergman:2014kza} and half a year later than \cite{Aganagic:2014oia}, we do not have any intention to claim the priority.  That said, our checks are largely independent of those that they performed, and can be considered as more pieces of evidence for their proposal. 

\section{Brane construction in 5d}\label{sec:brane}
\subsection{The web diagram for the $T_N$ theory}
\begin{figure}[t]
\begin{center}
\includegraphics[width=130mm]{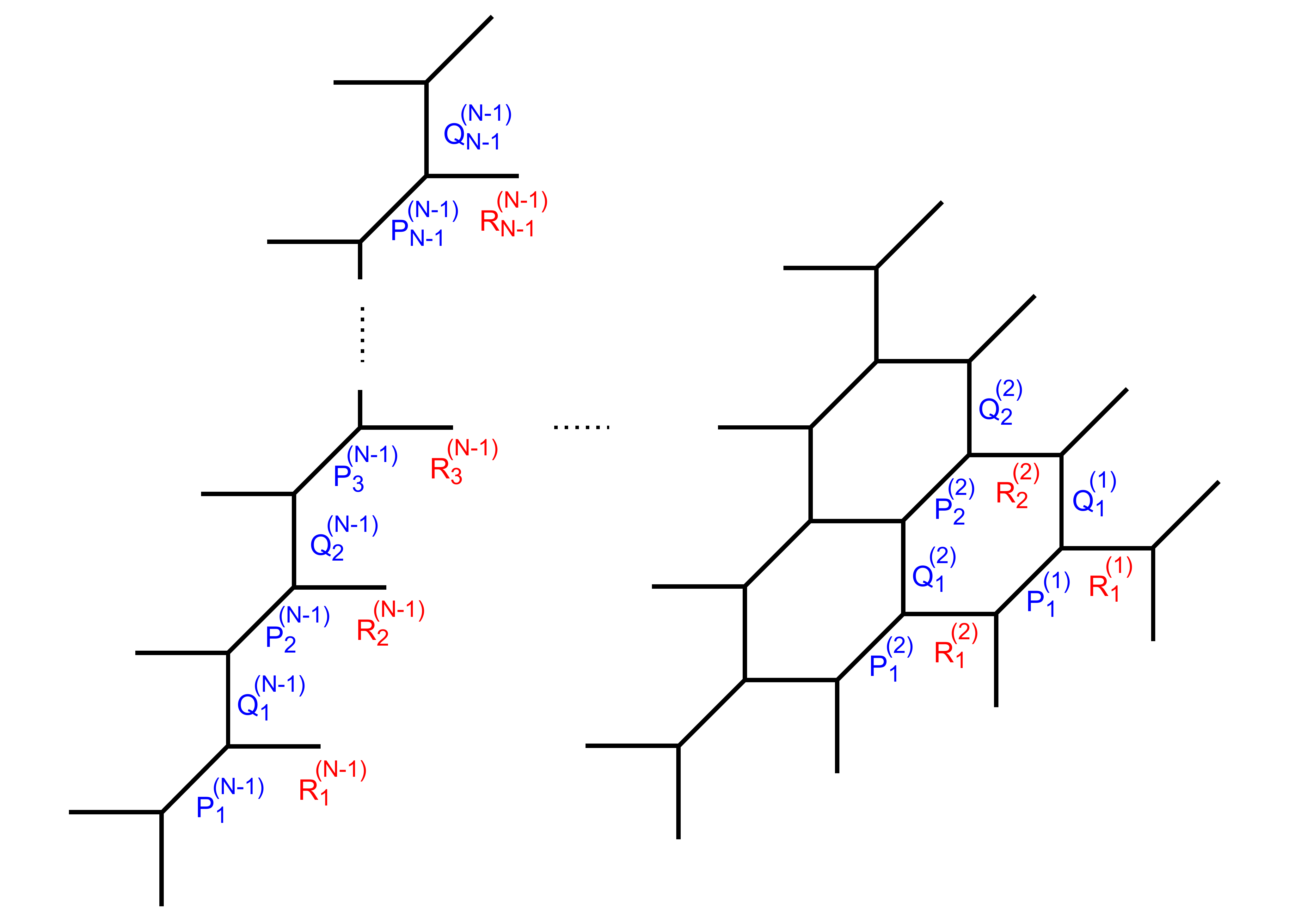}
\end{center}
\caption{The web diagram for the 5d $T_N$ theory. In our convention, a horizontal, vertical and diagonal line denotes a D5-brane, an NS5-brane and a $(1,1)$ 5-brane respectively. $P_k^{(n)}, Q_k^{(n)}, R_k^{(n)}, 1 \leq k \leq n$ for $1 \leq n \leq N{-}1$ are in a form $e^{iL}$ where $L$ is the length of the corresponding internal line.  They can be regarded as fugacities which appear in the computation of the partition function the 5d $T_N$ theory. }
\label{fig:TNweb}
\end{figure}

The $T_N$ theory  does not admit an obvious Lagrangian description, but the five-dimensional  version  can be explicitly realized in terms of a web of $(p, q)$ 5-branes \cite{Benini:2009gi}, shown in Fig.~\ref{fig:TNweb}.  It has  $N$ external D5-branes, $N$ external NS5-branes, and $N$ external $(1, 1)$ 5-branes, and they are connected with each other in the internal part of the diagram. The 5d $T_N$ theory lives on the intersection of the 5-branes. 
%%%%%%%%%%%%%%%%%%%%
%%%%%%%%%%%%%%%%%%%%

The global symmetry of the theory may be understood directly from the web diagram. For that, we put an orthogonal spacetime filling $(p, q)$ 7-brane on the end of each external $(p, q)$ 5-brane. This process does not break further supersymmetry. The lengths of the external 5-branes become finite and the global symmetry of the theory is realized on the $(p, q)$ 7-branes \cite{DeWolfe:1999hj}. In our case, we can end $N$ D5-branes on $N$ D7-branes, that gives  $\SU(N)$ symmetry.  The same is true for NS5-branes and (1,1) 5-branes. In total, we see that the theory realized by the web of Fig.~\ref{fig:TNweb} has the flavor symmetry $\SU(N) \times \SU(N) \times \SU(N)$.

\begin{figure}[t]
\begin{center}
\includegraphics[width=80mm]{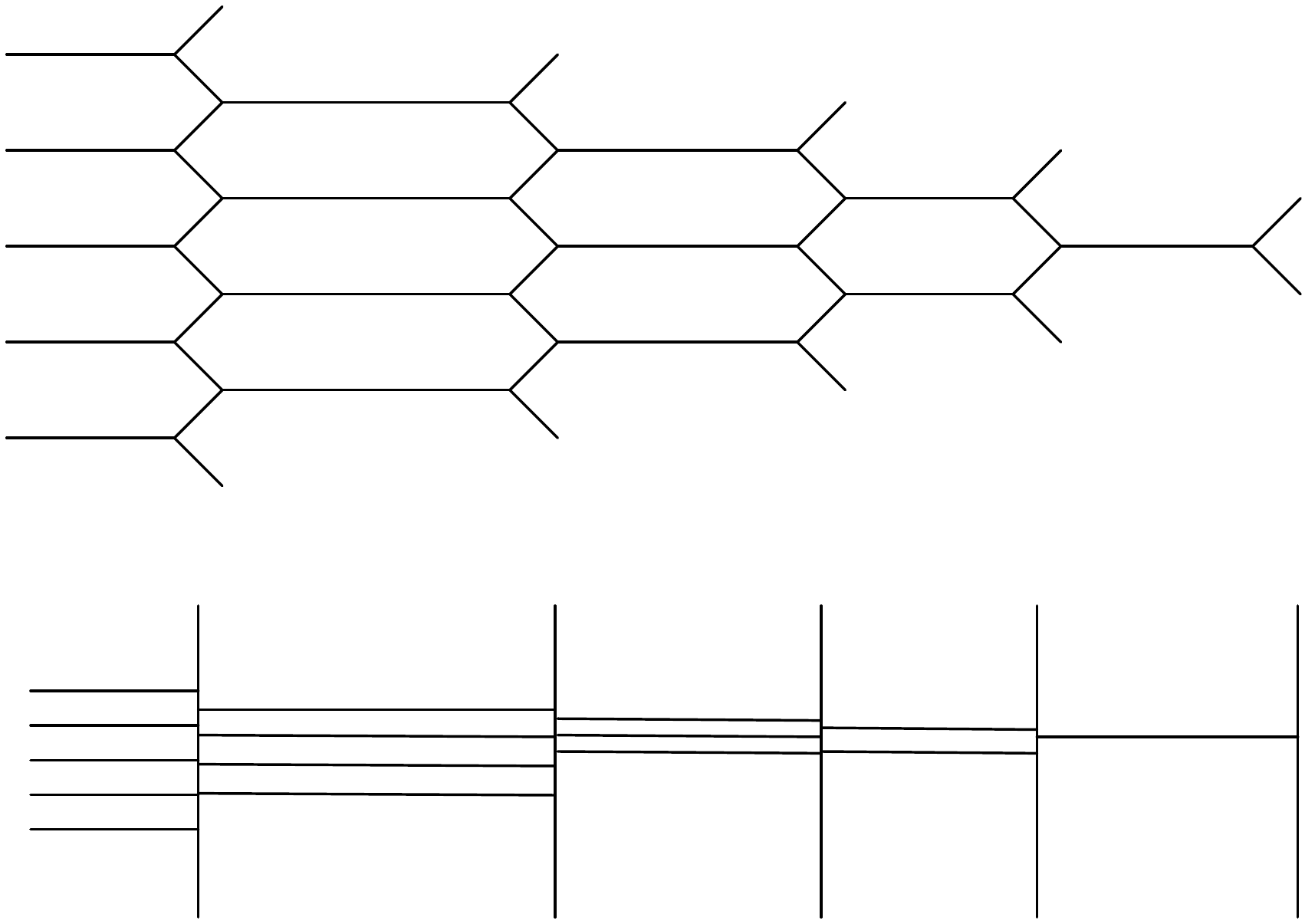}
\end{center}
\caption{Upper row: the web diagram when the masses for two $\SU(N)$s are equal and far larger than that for the third $\SU(N)$.  Lower row: the web diagram looks as this brane configuration, if one squints the eyes. The figures are shown for $N=5$.\label{fig:stupid}}
\end{figure}

At this point we can give a very crude argument relating the mass deformation of the $T_N$ theory and the linear quiver.  The mass terms for a single $\SU(N)$ correspond to the distance between $N$ parallel 5-branes. Let us give equal masses for two $\SU(N)$s so that they are far larger than the mass terms for the third $\SU(N)$.  The web diagram now becomes  the one shown in the upper row of Fig.~\ref{fig:stupid}.  This configuration looks very much like a simple brane configuration given in the lower row of the same figure, which realizes the linear quiver \begin{equation}
[\SU(N)]-\U(N{-}1)-\U(N{-}2)-\cdots-\U(2)-\U(1).
\end{equation} We see that the $\U(1)$ parts of the gauge groups are frozen,  since the two semi-infinite ends of a `vertical' brane in the lower figure is in fact not parallel, as one can see in the web diagram. Therefore the dynamical part of the linear quiver is \begin{equation}
[\SU(N)]-\SU(N{-}1)-\SU(N{-}2)-\cdots-\SU(2)-\SU(1).
\end{equation}
The objective of the rest of the section and of the paper is to make this rough argument more precise.

\subsection{The partition function}\label{sec:partpart}

\paragraph{Formalism.}
Given a  web of $(p, q)$ 5-branes, we can compute the exact partition function of the 5d theory compactified on a circle.  For that, we follow a chain of dualities, and view
the web diagram as the toric diagram of  a toric Calabi--Yau threefold \cite{Leung:1997tw}. In this picture, the 5d theory is realized as a low energy effective field theory of an M-theory compactification on this toric Calabi--Yau. For example, the web diagram corresponding to the 5d $T_N$ theory in Fig.~\ref{fig:TNweb} specifies a blow up of $\mathbb{C}^3/\mathbb{Z}_N \times \mathbb{Z}_N$ \cite{Benini:2009gi}. 

In this formulation, 5d BPS states  come from M2-branes wrapping various two-cycles inside the toric Calabi--Yau threefold \cite{Gopakumar:1998jq}, and their index  can be computed  by  the (refined) topological vertex \cite{Iqbal:2002we, Aganagic:2003db,Awata:2005fa, Iqbal:2007ii}, which can often be regarded as the 5d Nekrasov partition function of the corresponding 5d gauge theory \cite{Iqbal:2003ix, Iqbal:2003zz, Eguchi:2003sj, Taki:2007dh}. 

This is not the end of the story, however. The refined topological vertex computation in fact automatically contains the contribution of some BPS states that do not carry gauge charges and are decoupled from the 5d theory. 
Such  contributions come from strings between parallel external 5-branes (or M2-branes wrapping the corresponding two-cycles), and the web of $(p, q)$ 5-branes (or a toric diagram) allows us to easily identify them and strip them  \cite{Bergman:2013ala, Bao:2013pwa, Hayashi:2013qwa, Bergman:2013aca}.
These contributions  appear as  products of the plethystic exponentials, and we call them  decoupled factors.

\paragraph{Parametrization.}
Let us  now compute the partition function of the 5d  $T_N$ theory.  We assign parameters as shown  in Fig.~\ref{fig:TNweb}, but note that they satisfy \begin{equation}
P^{(k)}_aQ_a^{(k)} = Q_a^{(k+1)}P_{a+1}^{(k+1)}, \qquad R_a^{(k)} = R_1^{(k)}\left(P_1^{(k-1)}\cdots P_{a-1}^{(k-1)}\right)\left(P_2^{(k)} \cdots P_a^{(k)}\right)^{-1}.
\end{equation}
We parameterize them as follows. We first introduce $\lambda_{k; a}, a=1, \cdots, k$ for $k = 1, \cdots, N{-}2$ by
\begin{equation}
P_a^{(k)}Q_a^{(k)} = e^{-i\lambda_{k+1; a+1} + i\lambda_{k+1; a}},\label{para1}
\end{equation}
where 
\begin{equation}
\sum_{a=1}^{k+1}\lambda_{k+1;a} = 0. \label{Coulomb.sum}
\end{equation}
We then define  $m_{\text{bif, k}}, k = 1, \cdots, N{-}2$ by 
\begin{equation}
P_a^{(k)} = e^{i\lambda_{k+1; a} - i\lambda_{k; a} + im_{\text{bif}, k}}. \label{para2}
\end{equation}
with $\lambda_{1;1}=0$.  Next $m_{A, k}, k=1, \cdots, N$ are given by 
\begin{equation}
P_a^{(N{-}1)}Q_a^{(N{-}1)} = e^{-i\tilde{m}_{A, a+1}+i\tilde{m}_{A,a}}, \qquad P_a^{(N{-}1)} = e^{i\tilde{m}_{A, a} - i\lambda_{N{-}1; a}},
\end{equation}
for $a=1, \cdots, N{-}1$. Finally, the parameters $u_k$ for $k=1, \cdots, N{-}1$ are given by 
\begin{equation}
u_k = R_1^{(k)}Q_k^{(k) \frac{1}{2}}P_1^{(k)\frac{1}{2}}\left(P_2^{(k)}\cdots P_k^{(k)}\right)^{-\frac{1}{2}}\left(P_1^{(k-1)}\cdots P_{k-1}^{(k-1)}\right)^{\frac{1}{2}}. \label{para4}
\end{equation}
 We will later see that the parameters we defined through \eqref{para1}, \eqref{para2}--\eqref{para4} have a clear gauge theory interpretation. 

\paragraph{Explicit formulas.}
With the choice of the parameters \eqref{para1}, \eqref{para2}--\eqref{para4}, we apply the refined topological vertex to the web in Fig.~\ref{fig:TNweb}. 
In the computation, we choose the horizontal lines   to be the preferred directions.  
The calculation was performed in  \cite{Bao:2013pwa, Hayashi:2013qwa}. The quantity assigned to the web diagram is
\begin{equation}
\tilde{Z}_{T_N} =  Z_{\text{pert}} \cdot Z_{\text{inst}} \cdot Z_{\text{dec}}^{=},
\end{equation} and the genuine partition function $Z_{T_N}$ of $T_N$ is obtained by removing the decoupled factors:
\begin{equation}
Z_{T_N} = \left(M(t, q)M(q, t)\right)^{\frac{(N{-}1)(N{-}2)}{4}}\cdot \tilde{T}_{T_N}/(Z_{\text{dec}}^{=} \cdot Z_{\text{dec}}^{||} \cdot Z_{\text{dec}}^{//}). \label{ZTN}
\end{equation} 

Let us explain the ingredients in turn.
First, 
$t$ and $q$  are related to the $\Omega$--deformation parameters $(\epsilon_1, \epsilon_2)$ by $t=e^{i\epsilon_1}$ and $q=e^{-i\epsilon_2}$.
Then, $M(t, q)$ is the refined MacMahon function given by 
\begin{equation}
M(t, q) = \prod_{i,j=1}^{\infty}\left(1 - q^it^{j-1}\right)^{-1}.
\end{equation}
This factor comes from the perturbative contribution of the Cartan part of the vector multiplets. From the topological string point of view, it comes from the constant maps and cannot be captured by the refined topological vertex. We put it by hand in \eqref{ZTN} by adjusting its power by half of the dimension of the Coulomb branch moduli space of the $T_N$ theory.  Let us next give $Z_\text{pert}$ and $Z_\text{inst}$: 
{\small \begin{align}
Z_{\text{pert}} &= \prod_{i, j=1}^{\infty}\Big[\frac{\prod_{1\leq a \leq b \leq N{-}1}\left(1-e^{-i\lambda_{N{-}1;b}+i\tilde{m}_{A, a}}q^{i-\frac{1}{2}}t^{j-\frac{1}{2}}\right)\prod_{1\leq b < a \leq N}\left(1-e^{i\lambda_{N{-}1;b}-i\tilde{m}_{A, a}}q^{i-\frac{1}{2}}t^{j-\frac{1}{2}}\right)}{\prod_{k=2}^{N{-}1}\prod_{1\leq a < b \leq k}\left(1-e^{i\lambda_{k;a}-i\lambda_{k;b}}q^it^{j-1}\right)\left(1-e^{i\lambda_{k;a}-i\lambda_{k;b}}q^{i-1}t^{j}\right)}\nonumber\\
&\times \prod_{k=1}^{N{-}2}\prod_{1\leq a\leq b\leq k}\left(1-e^{i\lambda_{k+1;a}-i\lambda_{k;b}+im_{\text{bif}, k}}q^{i-\frac{1}{2}}t^{j-\frac{1}{2}}\right)\prod_{1\leq b < a \leq k+1}\left(1-e^{i\lambda_{k;b}-i\lambda_{k+1;a} - im_{\text{bif}, k}}q^{i-\frac{1}{2}}t^{j-\frac{1}{2}}\right)\Big],\nonumber\\ \label{TN.pert}
\\
Z_{\text{inst}}&= \sum_{\vec{Y}_1, \cdots, \vec{Y}_{N{-}1}}\prod_{k=1}^{N{-}1}u_k^{\sum_{a=1}^{k}|Y_{k; a}|}Z_{\text{inst}}\left(\vec{Y}_1, \cdots, \vec{Y}_{N{-}1}\right)\nonumber\\
&=\sum_{\vec{Y}_1, \cdots, \vec{Y}_{N{-}1}}\left\{\prod_{k=1}^{N{-}1}u_k^{\sum_{a=1}^{k}|Y_{k; a}|}z_{\text{vec}}(k)\right\}\left\{\prod_{k=1}^{N}z_{\text{fund}}(N{-}1; \tilde{m}_{A, k})\right\}\left\{\prod_{k=1}^{N{-}2}z_{\text{bif}}(k , k+1; m_{\text{bif}, k})\right\}.  \label{TN.inst}
\end{align}}

Here, 
 $\vec{Y}_k = \{Y_{k; a}\}, a=1 \cdots, k$ for $k=1, \cdots, N{-}1$ represent all possible Young diagrams. 
$z_{\text{bifund}}(k, l, m)$ is the contribution of  bifundamental hypermultiplets of gauge groups $\U(k) \times \U(l)$ with mass $m$ to the instanton partition function, and the explicit expression is 
{\small
\begin{equation}
z_{\text{bif}}(k, l ;m) = \prod_{a=1}^{k}\prod_{b=1}^{l}\prod_{s \in Y_{k; a}}\left[2i \sin\frac{E(k, l, a, b, s) - m + i\gamma_1}{2}\right]\prod_{\tilde{s} \in Y_{l; b}}\left[2i \sin\frac{E(l,k, b, a, \tilde{s}) + m + i\gamma_1}{2}\right], \label{bif.inst}
\end{equation}}
where the function $E(k, l, a, b, s)$ is defined as
\begin{equation}
E(k, l, a, b, s) = \lambda_{k; a} - \lambda_{l; b} + i(\gamma_1 + \gamma_2)l_{Y_{k; a}}(s) - i(\gamma_1 - \gamma_2)(a_{Y_{l; b}}(s) + 1).
\end{equation}
$\gamma_1$ and $\gamma_2$ are again related to the $\Omega$--deformation parameters by $i\epsilon_1=\gamma_1+\gamma_2$ and $i\epsilon_2=\gamma_1 - \gamma_2$. $l_Y(i, j)=Y_i - j$ and $a_Y(i, j)=Y_j^t - i$ denote some lengths inside the Young diagram $Y$ from a box specified by $(i, j)$. Here $Y_i$ denotes the height of the i--th column of a Young diagram $Y$, and $Y^t$ means the transpose of the Young diagram $Y$. $\lambda_{k; a}, a=1, \cdots, k$ are the Coulomb branch moduli of the $\U(k)$ gauge group. The other functions in \eqref{TN.inst} can be written by \eqref{bif.inst}: $z_{\text{vec}}(k)$ is the contribution of vector multiplets of a gauge group $\U(k)$ and it is 
\begin{equation}
z_{\text{vec}}(k) = \frac{1}{z_{\text{bif}}(k, k; i\gamma_1)}. \label{vec.inst}
\end{equation}
Also $z_{\text{fund}}(k; m)$ is the contribution from a fundamental hypermultiplet of a gauge group $\U(k)$ with mass $m$
\begin{equation}
z_{\text{fund}}(k; m) = z_{\text{bif}}(k, 0; m),\label{fund.inst}
\end{equation}
where the argument $0$ in \eqref{fund.inst} means that we do not have the product of $b$ and $Y_{l, b}$ in \eqref{bif.inst} and $\lambda_{0; b}=0$.

Finally,  the decoupled factors are 
\begin{align}
Z_{\text{dec}}^{=}& =\prod_{i, j=1}^{\infty} \prod_{1\leq a < b \leq N}\left(1-e^{i\tilde{m}_{A, a} - i\tilde{m}_{A, b}}q^i t^{j-1}\right)^{-1}\label{dec1}, \\
Z_{\text{dec}}^{||}&=\prod_{i, j = 1}^{\infty}\prod_{1 \leq a < b \leq N}\left(1 - \left(\prod_{k=a}^{b-1}u_ke^{\frac{i}{2}\left((k+1)m_{\text{bif, k}}-(k-1)m_{\text{bif, k-1}}\right)}\right)q^{i-1}t^{j}\right)^{-1}, \label{dec2} \\
Z_{\text{dec}}^{//} & = \prod_{i, j = 1}^{\infty}\prod_{1 \leq a < b \leq N}\left(1 - \left(\prod_{k=a}^{b-1} u_ke^{-\frac{i}{2}\left((k+1)m_{\text{bif, k}}-(k-1)m_{\text{bif, k-1}}\right)}\right)q^{i}t^{j-1}\right)^{-1}.
\label{dec3}
\end{align}
They are for  parallel external D5-branes,  NS5-branes, and 
 $(1, 1)$ 5-branes, respectively. We defined $m_{\text{bif}, N{-}1} = \frac{1}{N}\sum_{k=1}^N \tilde{m}_{A, k}$.

\paragraph{Interpretation.} 
From the explicit expression, it is now clear that the instanton part $Z_\text{inst}$ of the  partition function \eqref{TN.inst} is exactly that of the following linear quiver theory
\begin{equation}
[\SU(N)] - \U(N{-}1) - \U(N{-}2) - \cdots - \U(3)- \U(2) - \U(1), \label{lq1}
\end{equation} with all the Chern-Simons levels being zero.
But note that the sum of the Coulomb branch moduli of each gauge group is zero, because of \eqref{Coulomb.sum}. In \eqref{lq1}, the group in the square brackets $[\cdot]$ denotes a flavor symmetry. The parameter $u_k$ can be regarded as the instanton fugacity for the $\U(k)$ gauge group.

The partition function for the $T_N$ theory \eqref{ZTN} involves the division by decoupled factors from the  strings between the parallel external 5-branes in the $T_N$. In the case of the $T_3$ theory, the $E_6 \supset \SU(3) \times \SU(3) \times \SU(3)$ flavor symmetry was reproduced only after the removal of the decoupled factors \cite{Bao:2013pwa, Hayashi:2013qwa}. 
The decoupled factors \eqref{dec2} and \eqref{dec3} depend on the instanton fugacity, and therefore one cannot just say that the $T_N$ theory has the same partition function with  the quiver \eqref{lq1}. 
Instead we propose that \eqref{ZTN} yields the partition function of the following linear quiver theory
\begin{equation}
[\SU(N)] - \SU(N{-}1) - \SU(N{-}2) - \cdots - \SU(3) - \SU(2) - \SU(1). \label{lq2}
\end{equation}

\paragraph{On ``$\SU(1)$ instantons.''}
Let us discuss the physics of ``$\SU(1)$'' at the end of the quiver.   
As we recalled in the introduction,  the $T_2$ theory, which is  just the tri-fundamental of $\SU(2)_A\times \SU(2)_B\times\SU(2)_C$, flows to  the bifundamental of $\SU(2)_A\times \SU(1)$, under an appropriate choice of the mass terms. 
In this context, however, we can say even more.  The Nekrasov partition function of the ``$\SU(1)$ instantons'' coupled to two fundamentals, i.e.~the Nekrasov partition function of the $\U(1)$ theory after the removal of the decoupled factors, give back two fundamentals of $\SU(2)$.  

To see it, first consider the case $N=3$. 
The instanton part of the partition function \eqref{ZTN} reduces to
\begin{equation}
\sum_{Y_{2; 1}, Y_{2; 2}, Y_1}u_2^{|Y_{2;1}|+|Y_{2;2}|}u_1^{|Y_1|}Z_{\text{inst}}(\vec{Y}_2, \vec{Y}_1). \label{T3-1}
\end{equation}
Since the $T_3$ theory after the mass deformation can be thought of the $\SU(2)$ gauge theory with five flavors, \eqref{T3-1} should be  related to the partition function of the $\SU(2)$ gauge theory with five flavors if we redefine the parameters as \cite{Hayashi:2013qwa}
\begin{equation}
u_1 = e^{im_{f2}}, \quad u_2= u_2^{\prime}e^{-\frac{i}{2}m_{f2}}, \quad m_{\text{bif}, 1} = m_{f1}. \label{SU1}
\end{equation}
Here, $m_{f1}$ and $m_{f2}$ are the masses for the two fundamental hypermultiplets\footnote{The signs of the masses are different from the ones used in \cite{Hayashi:2013qwa}.  However, the instanton partition function of an $\SU(2)$ gauge theory is invariant under the flip of two signs of mass parameters for fundamental hypermultiplets since that is a part of the Weyl symmetry of the perturbative flavor symmetry.}, and $u_2^{\prime}$ is the instanton fugacity for the $\SU(2)$ gauge theory. 

Under this parameterization, we have 
\begin{equation}
\sum_{Y_1}e^{im_{f2}|Y_1|}Z_{\text{inst}}(\vec{Y}_1) = \prod_{i, j=1}^{\infty} \frac{\left(1 - e^{-i\lambda_{2;1}+im_{f2}}q^{i-\frac{1}{2}}t^{j-\frac{1}{2}}\right)\left(1 - e^{-i\lambda_{2;2}+im_{f2}}q^{i-\frac{1}{2}}t^{j-\frac{1}{2}}\right) }{\left(1 - e^{im_{f1}+im_{f2}}q^{i-1}t^{j}\right)\left(1 - e^{-im_{f1}+im_{f2}}q^{i}t^{j-1}\right)}. \label{pert.2f}
\end{equation} 
Then the factor with $(a, b)=(1, 2)$ in $Z^{||}$ \eqref{dec2} and the factor with $(a, b) = (1, 2)$ in $Z^{//}$ \eqref{dec3} cancel the denominator of \eqref{pert.2f}.
What remains is the numerator of \eqref{pert.2f}, which is the perturbative contribution of the fundamental hypermultiplet of the $\SU(2)$ with mass $m_{f2}$.  Combined with the perturbative contribution of the fundamental hypermultiplet with mass $m_{f1}$ already contained in \eqref{TN.pert}, we obtain the perturbative partition function of the following linear quiver theory 
\begin{equation}
 [\SU(3)] - \SU(2) - T_2, \label{lq3}
\end{equation}
 where $T_2$ simply denotes the two fundamental hypermultiplets with mass $m_{f1}$ and $m_{f2}$.

Since the part $Y_{2;1} =  Y_{2;1} = \emptyset$ in \eqref{T3-1} is now considered as the perturbative part, the genuine instanton part  is in fact 
\begin{equation}
\frac{\sum_{Y_{2; 1}, Y_{2; 2}, Y_1}u_2^{\prime |Y_{2;1}|+|Y_{2;2}|}e^{im_{f2}\left(|Y_1| - \frac{1}{2}(|Y_{2;1}|+|Y_{2;2}|)\right)}Z_{\text{inst}}(\vec{Y}_2, \vec{Y}_1)}{\sum_{Y_1}e^{im_{f2}|Y_1| }Z_{\text{inst}}(\emptyset, \vec{Y}_1)}.\label{T3-2}
\end{equation}
%By dividing \eqref{T3-2} by the new decoupling factor $Z_{\text{dec}}^{\prime}$ in the limit $u_k \rightarrow 0,\; k=3, \cdots, N{-}1$, 
The result becomes the $\SU(2) \cong \mathrm{Sp}(1)$ instanton partition function with five flavors \cite{Hayashi:2013qwa}. An important property for establishing the identification is that the expansion associated with $Y_1$ in \eqref{T3-2} stops at finite order with each Young diagram $Y_{2;1}$ and $Y_{2;2}$ fixed. This matches with the expectation from the field theory computation where the number of the terms involving the fugacity $e^{im_{f2}}$ is finite at each instanton number. % in the field theory computation for the $Sp(1)$ instanton partition function. 
We have checked this until $|Y_{2;1}|+|Y_{2;2}| = 3$ for several orders of $|Y_1|$.\footnote{Until the order we have checked, the summation of $Y_1$ stops at $|Y_1| = k$ if we consider $ |Y_{2;1}|+|Y_{2;2}| = k$.}

Let us now discuss the situation with general $N$. 
Since the termination of the summation of $Y_1$ in \eqref{T3-2} happens for each Young diagram $Y_{2;1}$ and $Y_{2;2}$, the termination should also occur in 
\begin{equation}
\frac{\sum_{\vec{Y}_{N{-}1}, \cdots, \vec{Y_2}, \vec{Y}_1}\prod_{k=3}^{N{-}1}u_k^{\sum_{a=1}^{k}|Y_{k; a}|}u_2^{\prime |Y_{2;1}|+|Y_{2;2}|}e^{im_{f2}\left(|Y_1| - \frac{1}{2}(|Y_{2;1}|+|Y_{2;2}|)\right)}Z_{\text{inst}}(\vec{Y}_{N{-}1}, \cdots, \vec{Y}_{3}, \vec{Y}_2, \vec{Y}_1)}{\sum_{Y_1}e^{im_{f2}|Y_1| }Z_{\text{inst}}(\emptyset, \cdots, \emptyset, \emptyset, \vec{Y}_1)}. \label{part.lq}
\end{equation}
for fixed instanton numbers of $u_2^{\prime}, u_k, \; k=3, \cdots, N{-}1$. This implies that the instanton partition function for the $\SU(1)$ part in \eqref{lq2} should have an interpretation as the fundamental hypermultiplet with the mass $m_{f2}$ coupled the $\SU(2)$ gauge fields. To see one of the evidence, one can check the flavor symmetry associated to $m_{f1}$ and $m_{f2}$.  Since the fundamental hypermultiplets couple to the $\SU(2)$, the flavor symmetry is enhanced to $\mathrm{SO}(4)$. We have checked that the instanton partition function \eqref{part.lq} is invariant under the Weyl symmetry of $\mathrm{SO}(4)$ for the $T_4$ theory until the order $\mathcal{O}(u_3u_2^{\prime})$.

\paragraph{Matching of the parameters.}
The argument so far strongly suggests that the partition function of the $T_N$ theory is exactly  the partition function of the linear quiver theory \eqref{lq2} or \eqref{lq3}. The global symmetry of the linear quiver theory is $\SU(N) \times \U(1)^{2N{-}2}$.  
The physical interpretation of the equality is that, when one gives generic mass terms for the two $\SU(N)$ flavor symmetries of the $T_N$ theory, the theory flows to the linear quiver theory \eqref{lq2}.

Let us study the relations between the mass parameters which break the $\SU(N) \times \SU(N)$ into $\U(1)^{2N{-}2}$ and the the parameters of the linear quiver theory. Each of the three $\SU(N)$ flavor symmetries is associated with the 7-branes attached to each parallel external 5-branes. Regarding the web diagram in Fig.~\ref{fig:TNweb}, all the 7-branes are separated from each other and the generic mass deformations are given to the three $\SU(N)$ global symmetries. The mass deformation is characterized by the length between the parallel external 5-branes. Let us first denote the mass deformations by $\mm_A$ for the parallel external D5-branes, $\mm_B$ for the parallel external NS5-branes, and $\mm_C$ for the parallel external $(1, 1)$ 5-branes.  
The three types of the mass deformations can be written by the fugacities that appear in Fig.~\ref{fig:TNweb}\footnote{The sign in front of $im_{C, k} - im_{C, k+1}$ in \eqref{SUN3} was chosen so that we deal with the three $\SU(N)$ flavor symmetries in a cyclically symmetric way.} 
\begin{eqnarray}
e^{im_{A,k} - im_{A,k+1}} &=& Q_k^{(N{-}1)}P_k^{(N{-}1)}, \label{SUN1} \\
e^{im_{B, k} - im_{B, k+1}} &=& P_1^{(k)}R_1^{(k)}, \label{SUN2}\\
e^{-(im_{C, k} - im_{C, k+1})} &=& R_k^{(k)}Q_k^{(k)}. \label{SUN3}
\end{eqnarray}
for $k=1, \cdots, N{-}1$. Then, combining \eqref{para1}, \eqref{para2}--\eqref{para4} with \eqref{SUN1}--\eqref{SUN3}, we obtain the relations 
\begin{align}
\mm_A &= \text{diag}\left(\tilde{m}_{A, 1}, \cdots, \tilde{m}_{A, N}\right) - m_{\text{bif}, N{-}1}\text{diag}\left(1, \cdots, 1 \right),\\
\mm_B &= \frac{1}{2}\left(m_{f1} + m_{f2}\right)H_1 + \sum_{k=2}^{N{-}1}\left(\frac{1}{2}m_{\text{bif}, k} + \frac{1}{k(k+1)}\sum_{a=2}^{k}\frac{8\pi^2 a}{g_a^2}\right)H_k,   \\
\mm_C &= \frac{1}{2}\left(m_{f1} - m_{f2}\right)H_1 +\sum_{k=2}^{N{-}1}\left(\frac{1}{2}m_{\text{bif}, k} - \frac{1}{k(k+1)}\sum_{a=2}^{k}\frac{8\pi^2 a}{g_a^2}\right)H_k.  
\end{align}
where we use the explicit form of the instanton fugacity $u_k = \exp\left(i\frac{8\pi^2}{g_k^2}\right)$. We introduced the notation
\begin{eqnarray}
H_k = \text{diag}\left(1, \cdots, 1, -k, 0, \cdots, 0 \right),
\end{eqnarray}
where there are $k$ entries of $1$ and $\tr H_k=0$.

\subsection{Higgsing one full puncture}

\begin{figure}[t]
\begin{center}
\includegraphics[width=60mm]{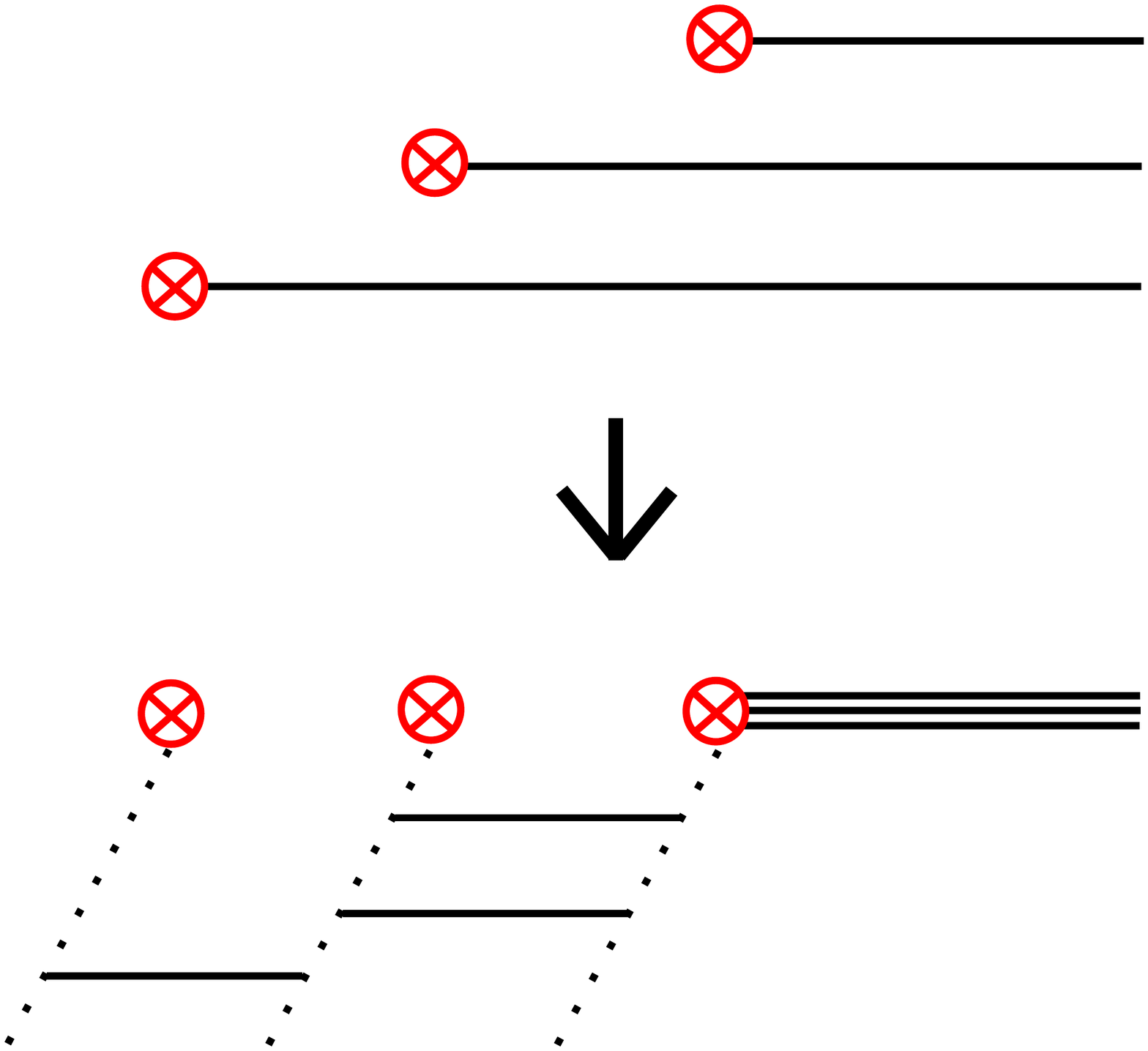}
\end{center}
\caption{The local deformation of 5-branes between 7-branes when we turn off some mass deformations. The red $\otimes$ represent a 7-brane. The dotted lines indicate another three directions where the 7-branes are extended.}
\label{fig:higgs}
\end{figure}

\paragraph{Brane construction.} 
By tuning the lengths of 5-branes, we can put several parallel external 5-branes on one 7-brane. Then the fractionated 5-branes can move between the 7-branes. An example of the process is depicted in Fig.~\ref{fig:higgs}.
%%%%%%%%%%%%%%%%%%%%
%%%%%%%%%%%%%%%%%%%%
This corresponds to the Higgs branch associated to one full puncture. 
 Suppose the fractionated 5-branes between the 7-branes are moved into infinity. Then, some 7-branes are disconnected from the web diagram and the number of the 7-branes gets reduced.
In the infrared  we obtain a different theory. In general, $n_i$ 5-branes may end on the $i$-th 7-brane where $\sum_i n_i = N$. Hence, different IR theories are classified by the partition $[n_i]$ of $N$. If the number of the bunches of $n_i$ 5-branes is $w_{n_i}$, the configuration carries a $\U(w_{n_i})$ global symmetry. The total global symmetry is then $\mathrm{S}\left(\prod_i \U(w_{n_i})\right)$ since the diagonal $\U(1)$ does not appear in the IR theory \cite{DeWolfe:1999hj}.
This datum $[n_i]$ is the same one introduced by Gaiotto  \cite{Gaiotto:2009we}.  In the case of the web diagram in Fig.~\ref{fig:TNweb}, the partition is given by $[1, 1, \cdots, 1]$ where we have $N$ 1's for all the three types of the parallel external 5-branes. Therefore, the web diagram corresponds to a sphere with three full punctures.  

We are interested in the partition function of this Higgsed system.  The general prescription to compute the partition function has already been presented in \cite{Hayashi:2014wfa}.  Let us illustrate this by considering a specific example, namely
 the case where $N{-}K$ upper parallel external D5-branes are put on one D7-brane. This corresponds to a puncture $[N{-}K, 1, \cdots, 1]$ where we have $K$ $1$'s. 
 
 The tunings can be described by \cite{Hayashi:2013qwa, Hayashi:2014wfa}
\begin{equation}
P_k^{(N{-}1)} = Q_k^{(N{-}1)} = \left(\frac{q}{t}\right)^{\frac{1}{2}}, \quad K+1 \leq k \leq N{-}1. \label{Higgs1}
\end{equation}
for the fugacities in Fig.~\ref{fig:TNweb}. Due to the constraints from the geometry of the web diagram. the tunings \eqref{Higgs1} induce another conditions for the fugacities in the interior of the web diagram
\begin{equation}
P^{(n)}_k = Q^{(n)}_k = \left(\frac{q}{t}\right)^{\frac{1}{2}}, \quad K+1 \leq k \leq n, \label{Higgs2}
\end{equation}
for $K+1 \leq n \leq N{-}2$. 

\paragraph{Perturbative part.}
Let us first look at the perturbative part \eqref{TN.pert}. By using the relations \eqref{Higgs1} and \eqref{Higgs2}, \eqref{TN.pert} reduces to
\begin{equation}
Z_{\text{pert}}= Z_{\text{pert},1} \cdot Z_{\text{pert}, 2} \cdot Z_{\text{pert}, 3}\cdot Z_{\text{singlets}},
\end{equation} where 
{\small \begin{align}
Z_{\text{pert},1}  &=  \prod_{i, j=1}^{\infty}\Big[\frac{\prod_{1\leq a \leq b \leq K}\left(1-e^{-i\lambda_{N{-}1;b}+i\tilde{m}_{A, a}}q^{i-\frac{1}{2}}t^{j-\frac{1}{2}}\right)\prod_{1\leq b < a \leq K}\left(1-e^{i\lambda_{N{-}1;b}-i\tilde{m}_{A, a}}q^{i-\frac{1}{2}}t^{j-\frac{1}{2}}\right)}{\prod_{k=K}^{N{-}1}\prod_{1\leq a < b \leq K}\left(1-e^{i\lambda_{k;a}-i\lambda_{k;b}}q^it^{j-1}\right)\left(1-e^{i\lambda_{k;a}-i\lambda_{k;b}}q^{i-1}t^{j}\right)}\nonumber\\
&\times\prod_{k=K}^{N{-}2}\prod_{1\leq a\leq b\leq K}\left(1-e^{i\lambda_{k+1;a}-i\lambda_{k;b}+im_{\text{bif}, k}}q^{i-\frac{1}{2}}t^{j-\frac{1}{2}}\right)\prod_{1\leq b < a \leq K}\left(1-e^{i\lambda_{k;b}-i\lambda_{k+1;a} - im_{\text{bif}, k}}q^{i-\frac{1}{2}}t^{j-\frac{1}{2}}\right)\Big],\nonumber\\ \label{TNHiggs.pert1}
\\
Z_{\text{pert},2}  &= \prod_{i, j=1}^{\infty}\prod_{1\leq b \leq K}\left(1 - e^{i\lambda_{K;b}-i\lambda_{K+1;K+1} - im_{\text{bif},K}}q^{i-\frac{1}{2}}t^{j-\frac{1}{2}}\right),\label{TNHiggs.pert2}\\
Z_{\text{pert},3}&= \prod_{i, j=1}^{\infty}\prod_{k=1}^{K-1}\Big[\frac{\prod_{1\leq a\leq b\leq k}\left(1-e^{i\lambda_{k+1;a}-i\lambda_{k;b}+im_{\text{bif}, k}}q^{i-\frac{1}{2}}t^{j-\frac{1}{2}}\right)}{\prod_{1\leq a < b \leq k}\left(1-e^{i\lambda_{k;a}-i\lambda_{k;b}}q^it^{j-1}\right)\left(1-e^{i\lambda_{k;a}-i\lambda_{k;b}}q^{i-1}t^{j}\right)}\nonumber\\
&\times\prod_{1\leq b < a \leq k+1}\left(1-e^{i\lambda_{k;b}-i\lambda_{k+1;a} - im_{\text{bif}, k}}q^{i-\frac{1}{2}}t^{j-\frac{1}{2}}\right)\Big], \label{TNHiggs.pert3}
\\
Z_{\text{singlets}}&= \prod_{i, j=1}^{\infty} \left[\prod_{k=K+2}^{N}\left(1 - q^it^{j-1}\right)^{k-K}\prod_{a \leq K,\; b \geq K+1}\left(1 - e^{-i\lambda_{N{-}1;b}+i\tilde{m}_{A,a}}q^{i-\frac{1}{2}}t^{j-\frac{1}{2}}\right)\right]. \label{TNHiggs.singlet}
\end{align}}

Here, $Z_{\text{pert,1,2,3}}$, \eqref{TNHiggs.pert1}--\eqref{TNHiggs.pert3}, are the perturbative partition function for the following linear quiver theory
\begin{equation}
[\SU(K)] - \SU(K) - \SU(K) - \cdots - \SU(K) - \SU(K-1) - \cdots - \SU(2) - \SU(1), \label{lq4}
\end{equation}
where an additional fundamental hypermultiplet whose contribution is \eqref{TNHiggs.pert2} is coupled to the rightmost  $\SU(K)$. The last factors \eqref{TNHiggs.singlet} are the contributions of singlet hypermultiplets that are decoupled from the linear quiver theory \eqref{lq4}. 

\paragraph{Instanton part.}
We then apply the conditions \eqref{Higgs1} and \eqref{Higgs2} to the instanton partition function \eqref{TN.inst}. First note that the tunings \eqref{Higgs1} and \eqref{Higgs2} trivialize some of the Young diagram summations. From the explicit form of \eqref{TN.inst} with \eqref{Higgs1} and \eqref{Higgs2}, we obtain a non-zero result only when $Y_{k, a} = \emptyset, K+1 \leq a \leq k$ for $K+1 \leq k \leq N{-}1$. In other words, when horizontal internal lines become external and end on 7-branes, the Young diagram summations associated to the horizontal lines are trivialized. Physically, since the horizontal lines become external and semi-infinite, the instanton particles corresponding to M2-branes wrapping the horizontal lines become infinitely heavy and non-dynamical.

With this simplification of the Young diagrams as well as the conditions \eqref{Higgs1} and \eqref{Higgs2}, there are various cancellations between the numerators and the denominators in the instanton partition function \eqref{TN.inst}. The final result is  
\begin{equation}
Z_{\text{inst}} = \sum_{\vec{Y}_1, \cdots, \vec{Y}_{N{-}1}}Z_{\text{inst}, 1} \cdot Z_{\text{inst}, 2} \cdot Z_{\text{inst}, 3}
\end{equation}
where
\begin{eqnarray}
Z_{\text{inst}, 1} &=& \left\{\prod_{k=K}^{N{-}1}u_k^{\sum_{a=1}^{K}|Y_{k; a}|}\tilde{z}^K_{\text{vec}}(k)\right\}\left\{\prod_{k=1}^{K}\tilde{z}^K_{\text{fund}}(N{-}1; \tilde{m}_{A, k})\right\}\left\{\prod_{k=K}^{N{-}2}\tilde{z}^K_{\text{bif}}(k , k+1; m_{\text{bif}, k})\right\},\nonumber\\  \label{TNHiggs.inst1}
\\
Z_{\text{inst}, 2} &=& \tilde{z}_{\text{fund}}^K\left(K; \lambda_{K-1;K-1} + m_{\text{bif,K}}\right), \label{TNHiggs.inst2}\\
Z_{\text{inst}, 3} &=& \prod_{k=1}^{K-1}u_k^{\sum_{a=1}^{k}|Y_{k; a}|}z_{\text{vec}}(k) \; z_{\text{bif}}(k , k+1; m_{\text{bif}, k}),  \label{TNHiggs.inst3}
\end{eqnarray}
where we defined
\begin{equation}
\tilde{z}^{K}_{\text{bif}}(k, l ;m) = \prod_{a=1}^{K}\prod_{b=1}^{K}\prod_{s \in Y_{k; a}}\left[2i \sin\frac{E(k, l, a, b, s) - m + i\gamma_1}{2}\right]\prod_{\tilde{s} \in Y_{l; b}}\left[2i \sin\frac{E(l,k, b, a, \tilde{s}) + m + i\gamma_1}{2}\right], 
\end{equation}
and similarly for $\tilde{z}^{K}_{\text{vec}}(k)$ and $\tilde{z}^{K}_{\text{fund}}(k; m)$ by \eqref{vec.inst} and \eqref{fund.inst} respectively with $z_{\text{bif}}(k, l ;m) $ replaced with $\tilde{z}^{K}_{\text{bif}}(k, l ;m)$. Therefore, we obtain the instanton partition function of the following linear quiver theory 
\begin{equation}
[\SU(K)] - \U(K) - \U(K) - \cdots - \U(K) - \U(K-1) - \cdots - \U(2) - \U(1), \label{lq5}
\end{equation}
where a fundamental hypermultiplet whose contribution is \eqref{TNHiggs.inst2} is coupled to the rightmost $\U(K)$ in \eqref{lq5}.

Now we need to remove the decoupled factors \eqref{dec2} and \eqref{dec3} with the conditions \eqref{Higgs1} and \eqref{Higgs2} inserted. We expect that the resulting instanton partition function becomes the one of the linear quiver \eqref{lq4}. 

The linear quiver theory can be also deduced from the web diagram that has one puncture of $[N{-}K, 1, \cdots, 1]$ type and two full punctures. 
%%%%%%%%%%%%%%%%%%%%
\begin{figure}[t]
\begin{center}
\includegraphics[width=100mm]{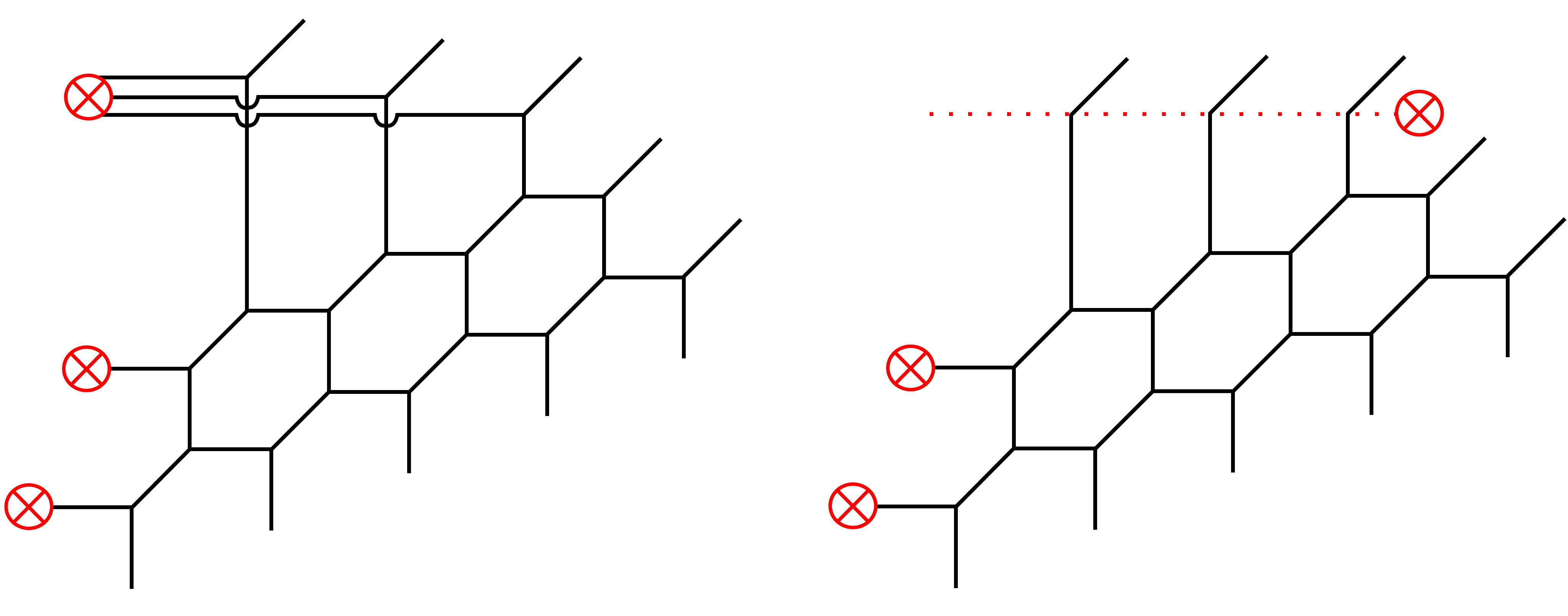}
\end{center}
\caption{The web diagram where the puncture associated with the parallel external D5-branes is $[3, 1, 1]$. The red dotted line denotes the branch cut for the D7-brane.}
\label{fig:T5Higgs}
\end{figure}
%%%%%%%%%%%%%%%%%%%%
Fig.~\ref{fig:T5Higgs} represents an example of such a web diagram. The original web diagram corresponds to the $T_5$ theory but now one of the punctures is Higgsed and three external D5-branes are put on one D7-brane. In analogy with Fig.~\ref{fig:stupid}, the web diagram on the right-hand side of Fig.~\ref{fig:T5Higgs} suggests the following linear quiver theory  
\begin{equation}
[\SU(2)] - \SU(2) - \SU(2) - \SU(2) - \SU(1),
\end{equation}
where an additional fundamental hypermultiplet is attached to the rightmost $\SU(2)$. The fundamental hypermultiplet is realized by strings between two D5-branes for that $\SU(2)$ and the D7-brane in the upper right part of the right figure of Fig.~\ref{fig:T5Higgs}. We can also see that the case with $[N{-}K, 1, \cdots, 1]$ and two full punctures can give rise to the linear quiver theory of \eqref{lq4} by using the corresponding web diagram.

\paragraph{General puncture.} 
We can see the general case by using a web diagram with a general puncture $[n_1, \cdots, n_k]$,  %with $n_1 \geq \cdots \geq n_k$ and 
$\sum_{i=1}^k n_i = N$, associated with the parallel external D5-brane. As in the right figure of Fig.~\ref{fig:T5Higgs}, we consider moving the D7-brane to which $n_i$ D5-branes are attached. Due to the brane annihilation, the $\SU(N-n)$ gauge theory reduces to the group 
$\SU\left(N-n - \sum_i H\left(n_i - n\right)\right)$. We introduced $H(x)$ that satisfies $H(x) = x$ for $x \geq 0$ and $H(x) = 0$ for $x < 0$. Also the D7-brane provides a fundamental hypermultiplet coupled to $\SU(N{-}n_i)$. Hence, the general case with $[n_1, \cdots, n_k]$ for one puncture yields a quiver theory 
\begin{equation}
\SU(v_1) - \cdots - \SU(v_n) \cdots - \SU(v_{N{-}1}), \label{general.web}
\end{equation}
where $v_n = N{-}n - \sum_{i=1}^k H(n_i - n)$ for $n=1, \cdots, N{-}1$. Furthermore, $w_{n}$ fundamental hypermultiplets are coupled to the $\SU(v_{n})$ where $w_n$ represents the number of $n$ appearing in $[n_1, \cdots, n_k]$. One can also check that $v_n$ satisfies \eqref{eq:generalR}. From the viewpoint of the quiver theory \eqref{general.web}, the global symmetry $\mathrm{S}\left(\prod_n \U(w_{n})\right)$ of the Higgsed $T_N$ theory is associated to the $w_n$ fundamental hypermultiplest coupled to the $\SU(v_n)$.

\section{Mass deformation from $T_N$ to $T_{N{-}1}$}\label{sec:rg}
In this section, we study the RG flow from the $T_{N}$ theory to the $T_{N{-}1}$ theory under a special form of mass deformations.

The masses of the $T_N$ theory take values in the Cartan subalgebra of $\SU(N)_{A} \times \SU(N)_B \times \SU(N)_C$. 
We denote them as $\mm_A, \mm_B$ and $\mm_C$ respectively.
Let us consider the following mass deformation;
\beq
\mm_A&=0, \nonumber \\
\mm_B&=\diag(m_{B}, \cdots, m_{B}, (1-N)m_{B} ), \nonumber \\
\mm_C&=\diag(m_{C}, \cdots, m_{C}, (1-N)m_{C} ).  \label{eq:specialmass}
\eeq
This mass deformation breaks the flavor symmetry as 
\beq
&\SU(N)_A \times \SU(N)_B \times \SU(N)_C  \nonumber \\
\to &\SU(N)_A \times \SU(N{-}1)_B \times \SU(N{-}1)_C \times \U(1)_B \times \U(1)_C.
\eeq

Furthermore, let us assume that the signs of the masses $m_B$ and $m_C$ are different, i.e., $m_Bm_C<0$. For concreteness we assume
\beq
m_B>0, ~~~m_C<0. \label{eq:masscondition}
\eeq
Then, under this mass deformation, we claim that the $T_N$ theory flows to the theory
\beq
[\SU(N)_{A}]- \SU(N{-}1)-T_{N{-}1} \label{eq:basicflow}.
\eeq
The meaning of this is as follows. The $\SU(N)_A$ is the flavor symmetry in the original theory,
and $\SU(N{-}1)$ is a gauge symmetry. There is a bifundamental of $\SU(N)_A \times \SU(N{-}1)$.
One of the three $\SU(N{-}1)$'s of the $T_{N{-}1}$ theory is gauged by the $\SU(N{-}1)$ gauge symmetry,
and the other two are matched to the $\SU(N{-}1)_B \times \SU(N{-}1)_C$ of the original UV theory. 

In the 5d version, we have zero Chern-Simons term for $\SU(N{-}1)$, and
the mass $m_{\rm bif}$ of the bifundamental of $\SU(N)_A \times \SU(N{-}1)$ and the gauge coupling $8\pi^2/g^2$ of the $\SU(N{-}1)$ are given by
\beq
m_{\rm bif}&=m_{B}+m_{C},\label{eq:bifmass} \\
\frac{8\pi^2}{g^2} &= \frac{N}{2} (m_B -m_C). \label{eq:instmass}
\eeq
Consistency checks on this proposal will be discussed below. 
There is a $\U(1)_{\rm bif}$ which rotates the bifundamental of $\SU(N)_A \times \SU(N{-}1)$, 
and another $\U(1)_{\rm inst}$ coming from the instanton current of $\SU(N{-}1)$.
These two $\U(1)$'s are matched to the $\U(1)_B \times \U(1)_C$ of the original theory in the way indicated by the above formulas 
for $m_{\rm bif}$ and $8\pi^2/g^2$.

We do not have many direct field-theoretical checks concerning the level of the Chern-Simons term.  The ones we have are: (i) Recursively applying this procedure as we will do in  Sec.~\ref{sec:qv}, we see that $\SU(N-1)$ has $2(N-2)$ fundamentals. In \cite{Intriligator:1997pq} it was shown that in this case we can only have zero Chern-Simons level to have nontrivial UV fixed point. (ii) If there is  nonzero Chern-Simons level $k$, the level $-k$ should also be possible. Then there should be two subtly-different versions of the 5d $T_N$ theory, but this seems not to be the case. 

\subsection{Matching of operators and states}
First, let us recall the following basic facts. We denote the $\U(1)_B \times \U(1)_C$ charges of operators or states as $q_B$ and $q_C$,
respectively. The Hilbert space is decomposed as 
\beq
\CH=\bigoplus_{q_B, q_C} \CH_{(q_B, q_C)},
\eeq
where $\CH_{(q_B, q_C)}$ is the subspace which has charge $(q_B, q_C)$.
By the BPS bound, the energies of the states in $\CH_{(q_B, q_C)}$ are bounded as
\beq
E \geq |q_B m_B+q_C m_C|~~~{\rm in~~} \CH_{(q_B, q_C)}. \label{eq:BPSbound}
\eeq
Then, operators charged under $\U(1)_B \times \U(1)_C$ can create only massive states.\footnote{By saying that an operator $O$ can create a particle, 
we mean that it has a nonvanishing matrix element $\bra{0 } O \ket{\rm particle} \neq 0$. 
Then the charges of the operator and the particle must be the same.  }

We consider the case $q_B =q_C$ and $q_B \neq q_C$ separately. This is because, according to \eqref{eq:instmass},
the states with $q_B \neq q_C$ have instanton charges in the IR theory.
The IR effective theory description is particularly good if $|m_{\rm bif}| \ll 8\pi^2/g^2$, i.e., the gauge coupling is small at the energy scale of $m_{\rm bif}$. 
Then the states with $q_B = q_C$ are light or massless, while the states with $q_B \neq  q_C$ are very heavy and involve instantons.

\subsubsection{Chiral operator matching for $q_B=q_C$}
The $T_N$ theory has the following 
chiral operators (in the language of 4d $\CN=1$ supersymmetry) which correspond to Higgs branch:
\beq
&(\mu_A)^{i_A}_{j_A}, ~(\mu_B)^{i_B}_{j_B},~(\mu_C)^{i_C}_{j_C}, \nonumber \\
&Q^{[i_{A,1},\cdots, i_{A,k}],[i_{B,1},\cdots, i_{B,k}],[i_{C,1},\cdots, i_{C,k}]}~(k=1,\cdots,N{-}1)
\eeq
where $i_A$, $i_B$, $i_C$ etc.~are the indices of $\SU(N)_A$, $\SU(N)_B$ and $\SU(N)_C$ respectively, 
and $[i_1 , \cdots, i_k]$ means that the indices are anti-symmetrized. 
The $\mu_{A,B,C}$ are in the adjoint representations of $\SU(N)_{A,B,C}$, and $Q$ are in the representation
$(\wedge^k, \wedge^k, \wedge^k)$ of $\SU(N)_A \times \SU(N)_B \times \SU(N)_C$, where $\wedge^k$
means the $k$-th anti-symmetric representation of $\SU(N)$. 

We denote the corresponding operators of the $T_{N{-}1}$ as $\mu'$ and $Q'$.
The bifundamental chiral operators of $\SU(N)_A \times \SU(N{-}1)$ are denoted as $B^{i_A}_{i_G}$ and $\tilde{B}^{i_G}_{i_A}$,
where $i_G$ is the index of the $\SU(N{-}1)$ gauge group.

By comparing flavor charges, it is easy to find the following operator matching between the UV and IR theories.
We often treat $B$ and $\tilde{B}$ as matrices. For the $\mu$ operators, 
\begin{align}
(\mu_A)^{i_A}_{j_A}&= (B \tilde{B})^{i_A}_{j_A} - \frac{1}{N} \delta^{i_A}_{j_A} \tr(B \tilde{B}), \label{eq:idmu1} \\
(\mu_X)^{i_X}_{j_X}&=(\mu'_X)^{i_X}_{j_X}~~(X=B,C~~i_X,j_X \leq N{-}1),  \label{eq:idmu2} \\
\frac{1}{2}[(\mu_B)^{N}_{N}+(\mu_C)^{N}_{N}]&=\frac{1}{N}\tr(B \tilde{B}),  \label{eq:idmu3}  \\
\frac{1}{2} [(\mu_B)^{N}_{N}-(\mu_C)^{N}_{N}] &\propto W_\alpha W^\alpha \label{eq:idmu4}
\end{align}
where $W_\alpha$ is the field strength superfield of the $\SU(N{-}1)$ gauge group. 
The last equation needs explanation. 
The $\mu$ operators are moment maps of the flavor symmetry, and they are in the same supermultiplets as the
flavor symmetry currents. Since the flavor current for the instanton charge is $F \wedge F$,
the corresponding operator with the correct mass dimension and flavor symmetry is $W_\alpha W^\alpha$.
Actually, in the Language of 4d $\CN=1$ supersymmetry, a moment map $\mu$ and a mass $m$
appear as $\int d^2\theta m \mu$. For the symmetry corresponding to $q_B -q_C$, we take $\mu \to W_\alpha W^\alpha$
and $m \to 1/g^2$.

For the $Q$ operators,
\beq
&Q^{[i_{A,1},\cdots, i_{A,k}],[i_{B,1},\cdots, i_{B,k}],[i_{C,1},\cdots, i_{C,k}]}  \nonumber \\
\sim &Q'^{[i_{G,1},\cdots, i_{G,k}],[i_{B,1},\cdots, i_{B,k}],[i_{C,1},\cdots, i_{C,k}]}B^{i_{A,1}}_{i_{G,1}} \cdots B^{i_{A,k}}_{i_{G,k}},  \label{eq:idQ1}
\eeq
and 
\beq
&Q^{[i_{A,1},\cdots, i_{A,k+1}],[i_{B,1},\cdots, i_{B,k},N],[i_{C,1},\cdots, i_{C,k},N]} \nonumber \\
\sim & \epsilon^{[i_{A,1},\cdots, i_{A,N}]} \epsilon_{[i_{G,1},\cdots, i_{G,N{-}1}]}Q'^{[i_{G,1},\cdots, i_{G,k}],[i_{B,1},\cdots, i_{B,k}],[i_{C,1},\cdots, i_{C,k}]}
\tilde{B}_{i_{A,k+2}}^{i_{G,k+1}} \cdots \tilde{B}_{i_{A,N}}^{i_{G,N{-}1}}, \label{eq:idQ2}
\eeq
where all the indices $i_B$ and $i_C$ are $ \leq N{-}1$. These are the simplest operator matchings one can think of.
For these equations to preserve the charges, the $\U(1)_B \times \U(1)_C$ charges of $B$ must be $(q_B,q_C)=(1,1)$.
This supports claim \eqref{eq:bifmass}.

\subsubsection{State matching for $q_B \neq q_C$}\label{sec:inststate}
In the $T_N$ theory, the chiral operators with charges $q_B \neq q_C$ are given by
\beq
&(\mu_B)^{i_B}_N,~~(\mu_B)_{i_B}^N,~~(\mu_C)^{i_C}_N, ~~(\mu_C)_{i_C}^N 
\eeq
which have charges $(q_B,q_C)= \pm (N,0)$ or $\pm (0,N)$, and 
\beq
&Q^{[i_{A,1},\cdots, i_{A,k}],[i_{B,1},\cdots, i_{B,k}],[i_{C,1},\cdots, i_{C,k-1}, N] },~~~
Q^{[i_{A,1},\cdots, i_{A,k}],[i_{B,1},\cdots, i_{B,k-1},N],[i_{C,1},\cdots, i_{C,k}] }
\eeq
which have charges $(q_B,q_C)=(k, k-N)$ or $(k-N,k)$ for $k=1, \cdots, N{-}1$.
Combining these results, the possible combinations of charges under $\U(1)_B \times \U(1)_C$ and the representation $r_A$ under 
$\SU(N)_A$ are given by
\beq
(q_B, q_C, r_A )= \pm (k, k-N, \wedge^k)~~~(k=0,1,\cdots, N),\label{eq:possiblecharge}
\eeq
where $ - \wedge^k$ formally means  $\wedge^{N{-}k}$. 
The cases $k=0$ and $N$ are given by $\mu$ operators,
while $1 \leq k \leq N{-}1$ are given by $Q$ operators. Tensor products of these representations are also possible by considering 
products of the operators.
Below, we reproduce these representations by performing the semi-classical quantization of instantons.

\paragraph{Semiclassical quantization of instantons.}
If we have an operator $O_{(q_B,q_C)}$ with charge $(q_B,q_C)$, we can consider states created by these operators,
\beq
O_{(q_B,q_C)}\ket{0} \in \CH_{(q_B, q_C)},
\eeq
which have the same charge as the operators. Their energies are bounded by \eqref{eq:BPSbound}.
The lowest mass states in each of $\CH_{(q_B, q_C)}$ may be BPS states.
We identify BPS states with charges \eqref{eq:possiblecharge} as the instanton particles of the $\SU(N{-}1)$ gauge group.
Instanton particles are obtained in semi-classical quantization by: 
(i) considering classical instanton solutions and (ii) quantizing the zero modes around the classical solutions.
The $\SU(N{-}1)$ gauge group is coupled to the bifundamental field $B, \tilde{B}$ and the $T_{N{-}1}$. The ``zero modes" of the $T_{N{-}1}$ are difficult to determine,
but they only affects the representations of the instanton particles under $\SU(N{-}1)_B \times \SU(N{-}1)_C$.
The representations under $\U(1)_B \times \U(1)_C \times \SU(N)_A$ can be obtained by quantization of zero modes of the bifundamental field. This is why we only consider the representation of $\U(1)_B \times \U(1)_C \times \SU(N)_A$ in \eqref{eq:possiblecharge}. We also do not discuss the gauge charge carried by the instanton particles; they will be affected by the ``zero modes'' of the $T_{N{-}1}$, and we assume that they are always canceled by appropriately combining various zero modes.

In a static instanton background, the action of the fermion $\psi$ in the hypermultiplet $B$ is given as
\beq
S=\int d^5x \left(  i\psi^\dagger_{i_A} \partial_{t} \psi^{i_A}- \bar{\psi}_{i_A} \gamma^i D_i \psi^{i_A} + m_{\rm bif}\bar{\psi}_{i_A} \psi^{i_A}   \right)
\eeq
where $i=1,2,3,4$ runs over spatial directions, and we have explicitly written the index $i_A$ of $\SU(N)_A$. 
We take $\gamma^i~(i=1,2,3,4)$ as the usual 4d gamma matrices, 
and take the gamma matrix in the time direction as $\gamma^t=-i\gamma^5$. Then $\bar{\psi}=\psi^\dagger \gamma^5$.

To perform semi-classical quantization of the zero modes, we assume $\psi$ has the form
\beq
\psi^{i_A} \simeq  a^{i_A}(t) \psi_0(x^i) ,
\eeq
where $\psi_0(x^i)$ is the zero mode of $\gamma^i D_i $ in the fundamental representation of $\SU(N{-}1)$,
and $a^{i_A}$ only depend on the time coordinate. The zero mode has a definite chirality $\gamma^5 \psi_k=\psi_k$, and hence
the action becomes
\beq
S=\int dt  \left( ia_{i_A}^\dagger \partial_t a^{i_A} +m_{\rm bif} a_{i_A}^\dagger a^{i_A}   \right).
\eeq
Canonical quantization gives
\beq
&\{ a^{i_A}, a_{j_A}^\dagger \}=\delta^{i_A}_{j_A}, ~~~\{a^{i_A}, a^{j_A}\}=0, \nonumber \\
&H = \frac{8\pi^2}{g^2} +  m_{\rm bif}(  a^{i_A}  a_{i_A}^\dagger  -\frac{N}{2}), \label{eq:instH}
\eeq
where $H$ is the Hamiltonian. We have included the classical energy $8\pi^2/g^2$ of the instanton particles.
The zero point energy $-N/2$ is required by the symmetry $a^{i_A} \leftrightarrow a_{i_A}^\dagger$, 
$m_{\rm bif} \leftrightarrow -m_{\rm bif}$.

Let $\ket{0}$ be the state with $a^\dagger_{i_A}\ket{0}=0$ for all $i_A$. Then, we obtain instanton particle states as
\beq
\ket{k}=a^{i_{A,1}} \cdots a^{i_{A,k}} \ket{0}.
\eeq
We denote the instanton charge as $q_{\rm inst}$ and the $\U(1)$ charge rotating the field $B$ as $q_{\rm bif}$.
Then the state $\ket{k}$ has the charge
\beq
(q_{\rm inst}, q_{\rm bif}, r_A)=(1, k-\frac{N}{2},  \wedge^k).
\eeq
This is in the representation $\wedge^k$ of $\SU(N)_A$, so we want to identify these states with the states \eqref{eq:possiblecharge}.
The case of the minus sign of $\pm$ in \eqref{eq:possiblecharge} is realized by anti-instantons.
This requires the identification of charges as
\beq
q_B=q_{\rm bif}+\frac{N}{2}q_{\rm inst},~~~q_C=q_{\rm bif}-\frac{N}{2}q_{\rm inst}.
\eeq

The Hamiltonian \eqref{eq:instH} gives the masses 
\beq
H=\frac{8\pi^2}{g^2} q_{\rm inst}+m_{\rm bif} q_{\rm bif}.
\eeq
This is equal to $m_B q_B+m_C q_C$ if and only if \eqref{eq:bifmass} and \eqref{eq:instmass} are satisfied.
This is the basis of our claim \eqref{eq:instmass}.

\subsection{Matching of the moduli space of vacua}
In this subsection, we compare the moduli space of vacua of the UV and IR theories.
We consider the case in which $\SU(N)_A$ also has the mass parameter of the form
\beq
\mm_A&=\diag(m_{A}, \cdots, m_{A}, (1-N)m_{A} ). 
\eeq
with
\beq
m_A+m_{\rm bif}=m_A+m_B+m_C=0. \label{eq:masssumzero}
\eeq
In this case, most of the $B, \tilde{B}$ fields in the IR theory become massless, and hence we get a larger Higgs branch.
The flavor symmetry is broken as $\SU(N)_A \to \SU(N{-}1)_A\times\U(1)_A$.

\subsubsection{IR theory}
In the IR theory \eqref{eq:basicflow}, the bifundamental of $\SU(N)_A\times \SU(N{-}1)$ splits into 
a bifundamental $b, \tilde{b}$ of $\SU(N{-}1)_A\times \SU(N{-}1)$ with mass $m_A+m_{\rm bif}=0$ and a fundamental of $\SU(N{-}1)$ 
with mass $(1-N)m_A+m_{\rm bif}$. We integrate out the massive fundamental, and get the quiver
\beq
[\SU(N{-}1)_A] - \SU(N{-}1)- T_{N{-}1}.
\eeq
All the fields in this quiver are massless.

This theory has a baryonic branch in which we give diagonal vevs to $b$ and $\tilde{b}$. 
In terms of gauge invariant operators, we define
\beq
\CB=b^{N{-}1},~~~\tilde{\CB}=\tilde{b}^{N{-}1},~~~\CM=\frac{1}{N{-}1} \tr b\tilde{b}.
\eeq
The baryonic branch is given as
\beq
\CB \tilde{\CB}=\CM^{N{-}1}.\label{eq:baryonb1}
\eeq
This is a hyperkahler manifold ${\mathbb C}^2/{\mathbb Z}_{N{-}1}$.
On this branch, the $\SU(N{-}1)$ gauge group is Higgsed, and the low energy theory consists of the $T_{N{-}1}$
theory and the neutral moduli fields \eqref{eq:baryonb1}. We will see that this result reproduces the moduli space of the UV $T_N$ theory deformed by the mass terms.

\subsubsection{UV theory}\label{sec:UV}
Now we study the moduli space of the $T_N$ theory. We use two different methods.
\paragraph{Field theory method.}
Generally in 4d $\CN=2$ or 5d $\CN=1$ theories, the potential on the Higgs branch under the mass deformation is \cite{AlvarezGaume:1983ab} 
\begin{equation}
\left|\sum_i m_i v_i \right|^2
\end{equation} where $v_i$ are the Killing vector of the $i$-th generator acting on the Higgs branch.
This can be easily seen in 4d from the fact that, in the language of 4d $\CN=1$ supersymmetry, the superpotential is given in terms of
the holomorphic moment maps $\mu_i$ as $\sum_i m_i \mu_i$, and the derivative of the holomorphic moment maps $\mu_i$ by moduli fields are given by
the holomorphic killing vectors $v_i$ by definition.
Therefore, after the mass deformation, 
we only have to keep operators uncharged under the Killing vector $\sum_i m_i v_i$ to see the moduli space of vacua.

The $T_N$ theory has operators $(\mu_{X})^{i_X}_{j_X}~(X=A,B,C) $ and $Q^{i_A i_B i_C}$ and $Q_{i_A i_B i_C}$.\footnote{
The vevs of operators $Q^{[i_{A,1},\cdots, i_{A,k}],[i_{B,1},\cdots, i_{B,k}],[i_{C,1},\cdots, i_{C,k}]}$ with $2 \leq k \leq N{-}2$ are 
determined by other operators' vevs and hence we do not have to consider them.}
When the relation \eqref{eq:masssumzero} is satisfied, the killing vector $\sum_i m_i v_i$ acts trivially on the operators $Q^{NNN}$ and $Q_{NNN}$.
It also acts trivially on $(\mu_{X})^N_N$.
Then we can give vevs to these operators.

The chiral ring relations of the $T_N$ theory are summarized in Appendix~\ref{sec:higgsrelations}. The relation $\tr \mu_A^k=\tr \mu_B^k=\tr \mu_C^k$ for any $k$, 
requires that the eigenvalues of the matrices $\mu_X~(X=A,B,C)$ are the same and we take their vevs as
\beq
\mu_X= -\frac{1}{N} \diag(\mu, \cdots, \mu, (1-N)\mu).
\eeq
The chiral ring relation \eqref{eq:qqmm} requires that they satisfy the relation\beq
Q^{NNN}Q_{NNN}=\mu^{N{-}1},   \label{eq:baryonb2}
\eeq
as was also discussed in \cite{Maruyoshi:2013hja}.

If we identify
\beq
Q_{NNN} \sim \CB,~~~Q^{NNN} \sim \tilde{\CB},~~~\mu \sim \CM,
\eeq
then \eqref{eq:baryonb2} is the same as \eqref{eq:baryonb1}. In fact, one can check that these identifications follow from the
operator matchings \eqref{eq:idQ1}, \eqref{eq:idQ2} and \eqref{eq:idmu3}.
Furthermore, by these vevs, the $T_N$ theory flows to the $T_{N{-}1}$ theory as discussed in \cite{Maruyoshi:2013hja}.
Therefore, the moduli spaces are matched between UV and IR description.

\paragraph{6d method.} 
Here we consider the $T_N$ theory in four dimensions. The 4d $T_N$ theory is realized by the compactification of the $\CN=(2,0)$ theory on
a Riemann sphere with three full punctures~\cite{Gaiotto:2009we}.
The Seiberg-Witten curve of the $T_N$ theory is given by
\beq
F_N(x,z)=x^N+ \sum_{k=2}^N \phi_k (z) x^{N{-}k}=0,
\eeq
where $z$ is a coordinate of the Riemann surface and $\phi_k(z)(dz)^k$ are $k$-th differential, i.e., sections of the $k$-th power of the canonical bundle
$K=T^*C$ of the Riemann surface.  Assuming that the punctures are at $z=z_X~(X=A,B,C)$, these $\phi_k$ are such that
the $N$ solutions of $x$ near these punctures are given by the eigenvalues of $\mm_X$,
\beq
x \sim \frac{\mm_X}{z-z_X}+{\rm lower~order}.
\eeq

Now, if the relation \eqref{eq:masssumzero} is satisfied, the curve can be factorized by tuning some Coulomb moduli as
\beq
F_N(x,z)=\left(x +(1-N) \phi_1(z) \right)  F_{N{-}1}(x+\phi_1(z),z),
\eeq
where $F_{N{-}1}(x,z)=x^{N{-}1}+\cdots$ is a curve of the $T_{N{-}1}$ theory, and $\phi_1$ is a one-form on the Riemann surface which has
poles at $z=z_X$ with residues $m_X$ ($X=A,B,C$), respectively. 
This is possible if and only if the sum of the residues 
is zero, i.e., $m_A+m_B+m_C=0$, since the sum of residues of any meromorphic one-form $\phi_1$ on a Riemann surface must be zero
due to the equation $\int_{C \setminus \{{\rm punctures} \} } d\phi_1=0$.

Intuitively, this branch is understood as follows. The theory may be realized by compactification of $N$ coincident M5 branes on the Riemann surface.
The above factorization corresponds to the case that one of the $N$ M5 branes is separated from the rest of the $N{-}1$ M5 branes.
See \cite{Xie:2014pua} for more systematic treatment.
If the curve is factorized in this way, we get quaternionic dimension one contribution to the Higgs branch from the motion of 
the separated one M5 brane as explained systematically in \cite{Yonekura:2013mya,Xie:2014pua} 
which generalize the earlier works \cite{Witten:1997sc,Hori:1997ab}. 
This should be identified with the baryon branch of \eqref{eq:baryonb1}.
Furthermore, it is clear that we get the $T_{N{-}1}$ theory with the curve $F_{N{-}1}$ on this branch. This is exactly as in the IR theory \eqref{eq:basicflow}.

\subsection{Strong coupling point and phase transition?}\label{sec:gcorrection}
When $m_{\rm bif} \neq 0$, it is possible to integrate out the bifundamental of $\SU(N)_{A}- \SU(N{-}1)$.
By doing that, we will be able to understand why  the condition \eqref{eq:masscondition} is necessary in five dimensions.

Suppose we have a simple gauge group $G$ and a hypermultiplet $H$ in some irreducible representation $r$ of $G$.
The gauge group $G$ has a coupling $g$ and the hypermultiplet $H$ has a mass $m$.
We would like to compute the low energy gauge coupling $g'$ after integrating out $H$.

By supersymmetry, we only need to compute it at the one-loop level.
This is because the gauge coupling is directly related to the masses of BPS instanton particles, and
masses of BPS particles are given by the central charge whose dependence on mass parameters is restricted. 
Another way of seeing the one-loop exactness is that if we extend the mass parameter to background vector superfield,
the correction to the gauge coupling is related by supersymmetry to Chern-Simons couplings of the form (gauge)$^2$(flavor).

The computation is straightforward, and 
we only write down the result. When we integrate out scalars or fermions or vectors in $d$-dimensions, 
the one-loop modification to the gauge coupling in spacetime dimension $d$ is given by
\beq
\frac{1}{g'^2}=\frac{1}{g^2} +C t_r \frac{\Gamma(2-d/2)}{(4\pi)^{d/2}} |m|^{d-4}
\eeq
where $t_r$ is the Dinkin index normalized to be $1/2$ for the fundamental representation of $\SU(N)$, and $C$ is given by
\begin{align}
C_s=&\frac{1}{3}  &&( {\rm complex~scalar}), \nonumber \\
C_f=&\frac{d_f}{3} &&({\rm complex~spinor}), \nonumber \\
C_v=& -\frac{26-d}{6}&&({\rm real~vector+ghost}),
\end{align} where $d_f$ is the complex dimension of the spinor.
One can check that this reproduces the usual result when $d= 4$.

A single hypermultiplet contains two complex scalars and fermions of dimension $d_f=4$.
By putting $d=5$ and using $\Gamma(-1/2)=-2 \sqrt{\pi}$, we get
\beq
\frac{8\pi^2}{g'^2}=\frac{8\pi^2}{g^2} - t_r  |m|. \label{eq:couplingcorr}
\eeq
This result would also be obtained by comparing the lowest mass states of instanton particles before and after integrating out 
the hypermultiplet $H$.

Now let us apply the above result to our case. By integrating out $B , \tilde{B}$, the $\SU(N{-}1)$ gauge coupling becomes
\beq
\frac{8\pi^2}{g'^2} &=\frac{8\pi^2}{g^2}- \frac{N}{2}|m_{\rm bif}| \nonumber \\
&= \frac{N}{2}\left[ (m_B -m_C) - |m_B+m_C| \right]. 
\eeq
This is positive as long as the condition \eqref{eq:masscondition} is satisfied. However, 
when one of the masses, say $m_B$, goes to zero, the coupling becomes infinitely large.
In that case, the description in terms of the effective theory \eqref{eq:basicflow} breaks down and 
the $T_N$ theory with $m_B=0$ should flow to some strongly coupled theory.
Note that when $m_B=0$, the symmetry should be restored to $\SU(N)_B$, thus it is {\it a priori} expected that something must happen at this point.

If we further take $m_B$ to be negative, the IR coupling formally becomes negative. So there must be some phase transition at $m_B=0$.
It would be very interesting to study this phase transition and the theory at $m_B<0$.

\section{Mass deformation to linear quivers}\label{sec:qv}

In the last section, we argued that the $T_N$ theory, when deformed by mass terms breaking $\SU(N)_B$ to $\SU(N{-}1)\times \U(1)$ and similarly for $\SU(N)_C$, becomes in the IR the theory of the form $[\SU(N)_{A}]- \SU(N{-}1)-T_{N{-}1}.$
In this section, we study what happens when we give generic mass terms to $\SU(N)_{B,C}$.
We also study the case when we replace the full puncture for $\SU(N)_A$ by more general ones. 

\subsection{Recursive application of the basic deformation}
To analyze the effect of generic mass terms,
we can first give the mass terms preserving $\SU(N{-}1)_{B,C}$, and then
 add masses which break $\SU(N{-}1)_{B,C}$.
For convenience, we define generators of Cartan subalgebra of $\SU(N)$ as
\beq
H_k = \diag(1,\cdots,1, -k,0,\cdots,0)~~~(k=1,2,\cdots,N{-}1)
\eeq
where there are $k$ entries of $1$ so that $\tr H_k=0$. Then, consider the mass matrices
\beq
\mm_B= m_{B,N{-}1}H_{N{-}1}+m_{B,N{-}2}H_{N{-}2} ,\nonumber \\
\mm_C= m_{C,N{-}1}H_{N{-}1}+m_{C,N{-}2}H_{N{-}2}.
\eeq
When $m_{B,N{-}2}=m_{C,N{-}2}=0$, these mass matrices are reduced to the previous ones with $m_B=m_{B,N{-}1}$ and  $m_C=m_{C,N{-}1}$.

If $m_{X,N{-}2}~(X=B,C)$ are much smaller than $m_{X,N{-}1}$, we first obtain the theory \eqref{eq:basicflow},
\beq
[\SU(N)_{A}]- \SU(N{-}1)-T_{N{-}1} . \label{eq:flowstep1}
\eeq
The $\SU(N{-}1)_B \times \SU(N{-}1)_C$ of the $T_{N{-}1}$ is now deformed by the masses $m_{X,N{-}2}$, and hence
by using our proposal again to $T_{N{-}1}$, we get
\beq
[\SU(N)_{A}]- \SU(N{-}1)-\SU(N{-}2)-T_{N{-}2} . \label{eq:flowstep2}
\eeq 
We denote the gauge couplings of $\SU(N{-}1)$ and $\SU(N{-}2)$ as $g_{N{-}1}$ and $g_{N{-}2}$, respectively.

There is one point one should be careful about. When the theory flows from \eqref{eq:flowstep1} to \eqref{eq:flowstep2},
some of the heavy ``fields" in the $T_{N{-}1}$ theory are integrated out. We have seen in the previous subsection~\ref{sec:gcorrection} that
the coupling constants receive quantum corrections when massive fields are integrated out. 
However, because the $T_{N{-}1}$ is non-Lagrangian,
we cannot perform the one-loop calculation to determine the corrections.

Here we simply write down the correction which are consistent with other analysis we performed. We denote the couplings of $\SU(N{-}1)$ in \eqref{eq:flowstep1} and \eqref{eq:flowstep2}
as $g_{N{-}1,b}$ and $g_{N{-}1,a}$, respectively. The coupling $g_{N{-}1,b}$ of \eqref{eq:flowstep1} is just given by \eqref{eq:instmass}, and
the difference $g_{N{-}1,a}^{-2}-g_{N{-}1,b}^{-2}$ is expected to depend only on $m_{X,N{-}2}~(X=B,C)$. It is given by
\beq
\frac{8\pi^2}{g^2_{N{-}1,a}}-\frac{8\pi^2}{g^2_{N{-}1,b}}= -\frac{N{-}2}{2} (m_{B,N{-}2}-m_{C,N{-}2}). \label{eq:TNout}
\eeq

The choice of the coefficient $(N-2)/2$ can be explained as follows. 
Let us consider an instanton of the $\SU(N{-}1)$ gauge group.
This gauge group is coupled to the bifundamentals of $\SU(N)_A \times \SU(N{-}1)$ and $\SU(N{-}1) \times  \SU(N{-}2)$, 
and we can perform semi-classical quantization as in subsection~\ref{sec:inststate}. The mass spectrum is given by
\beq
m_{\rm inst} &=\frac{8\pi^2}{g^2_{N{-}1,a}}+(k-\frac{N}{2})m_{{\rm bif},N{-}1} -(k'- \frac{N{-}2}{2}) m_{{\rm bif},N{-}2} \nonumber \\
&=\left[k m_{B,N{-}1} +(k-N) m_{C,N{-}1} \right] - \left[k' m_{B,N{-}1} +(k'-N+2) m_{C,N{-}1}   \right],
\eeq
where $m_{{\rm bif},N{-}1}=m_{B,N{-}1}+m_{C,N{-}1}$ and $m_{{\rm bif},N{-}2}=m_{B,N{-}2}+m_{C,N{-}2}$ are bifundamental masses, and
$k=0,\cdots,N$ and $k'=0,\cdots,N{-}2$.
For example, the state $k=k'=0$ has $m_{\rm inst}=-Nm_{C,N{-}1}+(N{-}2)m_{C,N{-}2}$. This is the same as the BPS mass
created by the operator $(\mu_C)^N_{N{-}1}$. For more detailed comparison, it is necessary to determine which instanton states 
do or do not have gauge charges.  This can be done by putting the theory on $S^4$ times the time direction and compute the index. This type of analysis was performed in \cite{Bergman:2014kza}. Here we preferred to present a more elementary semi-classical analysis, that does not require the full machinery of the index computation. 

Repeating the above procedure, we get the following result. We give masses of the form
\beq
\mm_B= \sum_{k=1}^{N{-}1} m_{B,k} H_k,~~~\mm_C= \sum_{k=1}^{N{-}1} m_{C,k} H_k.\label{eq:generalmass}
\eeq
Then, we get a linear quiver
\beq
[\SU(N)_A] - \SU(N{-}1)- \cdots-\SU(2)-T_2,
\eeq
where $T_2$ is just two fundamental hypermultiplets. Let $m_{{\rm bif},k}$ be the mass of the bifundamental of $\SU(k+1)\times \SU(k)$,
$m_{f1}$ and $m_{f2}$ the masses of the fundamentals in $T_2$, and $g_k$ the gauge coupling of $\SU(k)$.
Then, we get
\beq
m_{{\rm bif}, k} &=m_{B,k}+m_{C,k},~~~ (k=2,\cdots,N{-}1)   \\
m_{f1} &=m_{B,1}+m_{C,1},  ~~~~~ m_{f2} =m_{B,1}-m_{C,1}, \\
\frac{8\pi^2}{g^2_k} &=\frac{k+1}{2} (m_{B,k}-m_{C,k})-\frac{k-1}{2} (m_{B,k-1}-m_{C,k-1})~~(k=3, \cdots,N{-}1), \\
\frac{8\pi^2}{g^2_2} &= \frac{3}{2} (m_{B,2}-m_{C,2}).
\eeq
This is in complete agreement with the result obtained in Sec.~\ref{sec:brane}.

Note that if we formally go one step further, we get
\beq
[\SU(N)_A] - \SU(N{-}1)- \cdots-\SU(2)-``\SU(1)". \label{eq:originallinearq}
\eeq
and the couplings
\beq
 \frac{8\pi^2}{ g_2^2}&= \frac{3}{2} (m_{B,2}-m_{C,2})-\frac{1}{2} (m_{B,1}-m_{C,1}), \label{eq:modg2}\\
 \frac{8\pi^2}{ g_1^2}&= (m_{B,1}-m_{C,1}). \label{eq:modg1}
\eeq
This has the following interpretation. By using the formula \eqref{eq:couplingcorr}, we can see that the coupling \eqref{eq:modg2} is
precisely the one obtained by integrating out the fundamental with mass $m_{f2}$ in $T_2$.
This is the $T_2$ version of what has happened in \eqref{eq:TNout}.
Furthermore, the ``$\SU(1)$" coupling $8\pi^2/g^2_1$ is equal to $m_{f2}$. So we may formally think of this fundamental
as the ``$\SU(1)$ instanton".   We discussed that indeed, this fundamental hypermultiplet comes from the instanton of $\U(1)$ in the computation of partition functions in Sec.~\ref{sec:partpart}.

Next let us consider the case when we replace the full puncture giving $\SU(N)_A$ with a more general puncture of type $Y=[n_1,\ldots,n_p]$, with $\sum n_i=N$. 
This can be realized by giving a vev to $\mu_A$ of the $T_N$ theory of the form \begin{equation}
\mu_A=J_{n_1}\oplus \cdots \oplus J_{n_p} \label{eq:nilpvev}
\end{equation} where $J_{n}$ is the nilpotent Jordan block of size $n$. 

After the general mass deformation \eqref{eq:generalmass}, we have the linear quiver \eqref{eq:originallinearq}, where $\mu_A$ is given by the quadratic combination of the leftmost bifundamental.  The effect of  a nilpotent vev of the form \eqref{eq:nilpvev} to $\mu_A$ to a linear quiver whose gauge groups are all $\SU(N)$ was studied in detail in Sec.~12.5 of \cite{Tachikawa:2013kta}. There, it was shown that the rank of the $i$-th gauge group from the left is reduced by $\rank (\mu_A)^{i}$, and the additional hypermultiplets in the fundamental of the $i$-th gauge group is given by the number of times $i$ appears in 
$[n_1,\ldots,n_p]$. 
The same argument can be applied verbatim when we start from the quiver \eqref{eq:originallinearq}.

Therefore, the resulting linear quiver is of the form  \begin{equation}
\SU(v_{1})-\SU(v_{2})-\cdots -\SU(v_{N{-}2})-\SU(v_{N{-}1})\label{eq:generalquiver}
\end{equation} with additional $w_i$ fundamental hypermultiplets for $\SU(v_i)$, where
$w_k$ is the number of times $k$ appears in the partition $Y=[n_i]$, and $v_i$ are defined by the relation \begin{equation}
v_{N{-}1}=1,~v_N:=0;\quad  2v_i=v_{i-1}+v_{i+1} + w_i\ \text{for $i=2,\ldots,N{-}1$},
\end{equation}  so that every node has zero beta function when considered as a 4d gauge group. 
Let $K=N{-}n_1$.  We have  $v_{N{-}i}=i$ for $i \leq K$,  since $w_i =0 $ for $i > N{-}K=n_1$.
Then $v_{N{-}K}=K$, and after that the gauge groups are decreasing as $K \geq v_{N{-}K-1} \geq \cdots \geq v_1$.
%This linear quiver  has the same form as the 3d linear quiver description of the 3d theory $T_Y(\SU(N))$, introduced in \cite{Gaiotto:2008ak}. 

\subsection{Seiberg-Witten curves in 4d}\label{sec:curves}
Here we derive the Seiberg-Witten curve of linear quiver gauge theory from the curve of the mass-deformed $T_N$ theory.
More generally, we consider a theory realized by a Riemann sphere with two full punctures and one arbitrary puncture $Y$.
Then we introduce generic masses to the $\SU(N)_B \times \SU(N)_C$ flavor symmetry of the full punctures.

\begin{figure}
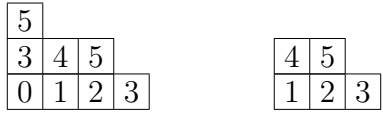

\[
\young(5,345,0123)
\qquad\qquad
\young(45,123)
\]
\caption{Left: Young diagram for $Y=[3,2,2,1]$. The numbers inside the boxes are the $p_k$ defined in \eqref{eq:defpk}. Right: Removing the
leftmost column. The numbers inside the boxes represent the $\ell=1,2,\cdots, K$.}\label{Young}
\end{figure}

Let $z$ be a coordinate of the Riemann sphere.
We put the full punctures at $z = \pm 1$ and the puncture $Y$ at $z=0$. The curve is given by
\beq
x^N+\sum_{k=2}^N \phi_k(z) x^{N{-}k}=0 , \label{eq:Ycurve}
\eeq
where the $k$-th differential $\phi_k$ is given by
\beq
\phi_k=\frac{1}{(1-z)^{k-1}(1+z)^{k-1}}\left( \frac{2^{k-1}M_{B,k}}{z(1-z)}+\frac{(-2)^{k-1}M_{C,k}}{ z(1+z)} + \frac{u^{k}_2}{z^2}+\cdots+\frac{u^{k}_{p_k}}{z^{p_k}} \right).
\eeq
Note that $\phi_k (dz)^k$ are finite at $z=\infty$ so that there is no puncture at $z =\infty$. 
The $M_{B,k}$ and $M_{C,k}$ are related to the mass parameters of $\SU(N)_B$ and $\SU(N)_C$ as
\beq
 \det( x -\mm_{X})= x^N+\sum_{k=2}^N (-1)^kM_{X,k} x^{N{-}k}~~~(X=B,C).
\eeq
The $p_k$ are the numbers associated to $Y$ explained by Gaiotto \cite{Gaiotto:2009we}.
Explicitly, if $Y$ is given by a partition of $N$ as $Y=[n_1,n_2,....]$ ($n_1 \geq n_2 \geq \cdots$)
and $Y^t=[n'_1,n'_2,\cdots]$ is its dual obtained by transposing the Young diagram of $Y$, $p_k$  is given by
\beq
p_k=k - a~~~(n'_1+\cdots+n'_{a-1} <k \leq n'_1+\cdots+ n'_a). \label{eq:defpk}
\eeq
In terms of the Young diagram of $Y$, $a$ is the height of the $k$-th box counted from left to right and bottom to top.
See the left of Fig.~\ref{Young}.

Now we take the limit $z \to 0$. Retaining only the most significant terms, we get
\beq
\phi_k \to  c_k+ \frac{\mu_k}{z} + \frac{u^{k}_2}{z^2}+\cdots+\frac{u^{k}_{p_k}}{z^{p_k}} , \label{eq:limsing}
\eeq
where $c_k=2^{k-1}(M_{B,k}+(-1)^{k}M_{C,k})$ and $\mu_k=2^{k-1}(M_{B,k}-(-1)^{k}M_{C,k})$. More precisely, our scaling limit is
\beq
z \sim \epsilon, ~~x \sim \epsilon^{-1},~~c_k \sim  \epsilon^{-k},~~\mu_k \sim  \epsilon^{-k+1},~~u^k_\ell \sim  \epsilon^{-k+\ell},
\eeq
and then take $\epsilon \to 0$. The scaling of $x$ is determined so that the Seiberg-Witten differential $\lambda=xdz$ is fixed.

Let $K=p_N=N{-}n_1$ as before. Also, define $q_\ell$ such that 
\beq
\ell \leq p_k \Leftrightarrow k \geq q_\ell.
\eeq
The explicit form of $q_\ell$ will be obtained later. Then, \eqref{eq:Ycurve} becomes
\beq
0=z^K\left(x^N+\sum_{k=2}^N c_k x^{N{-}k} \right)+z^{K-1}\left(\sum_{k=2}^N \mu_k x^{N{-}k} \right) 
+\sum_{\ell =2}^K z^{K-\ell} \left(  \sum_{k=q_\ell}^N u^k_\ell x^{N{-}k}  \right).
\eeq
Defining
\beq
\psi'_1=\frac{\sum_{k=2}^N \mu_k x^{N{-}k} }{x^N+\sum_{k=2}^N c_k x^{N{-}k}}, ~~~~~
\psi'_\ell=\frac{ \sum_{k=q_\ell}^N u^k_\ell x^{N{-}k} }{x^N+\sum_{k=2}^N c_k x^{N{-}k}},
\eeq
we get the curve
\beq
0=z^K+\psi'_1z^{K-1}+\sum_{\ell=2}^K \psi'_\ell z^{K-\ell}. \label{eq:dualYcurve}
\eeq
Furthermore, the Seiberg-Witten differential is $\lambda =xdz \cong -zdx$ up to a total derivative term.
Therefore, we get a curve of a class S theory of $A_{K-1}$ type by changing the roles of $z$ and $x$.

There are $N$ simple punctures at the solutions of $x^N+\sum_{k=2}^N c_k x^{N{-}k} =0 $. The parameters $c_k$ are related to the positions of
the simple punctures, and $\mu_k$ are related to the mass parameters of these simple punctures. 
From the results of the previous sections, the condition that the bifundamentals are massless is given by $\mm_B+\mm_C=0$.
In that case, we get $\mu_k=2^{k-1}(M_{B,k}-(-1)^{k}M_{C,k})=0$, consistent with the fact that $\mu_k$ are related to the mass parameters
at the simple punctures.

We also have a puncture at $x=\infty$.
Let $x'=1/x$. Then we get
\beq
\psi'_1dx \sim dx' ,~~~~~\psi'_\ell (dx)^\ell \sim \frac{(dx')^\ell}{x'^{2\ell-q_\ell}}.
\eeq
Therefore, this puncture, which we denote as $Y'$, is given by singularities of order $p'_\ell=2\ell-q_\ell$.

It turns out that $Y'$ is obtained from the Young diagram $Y$ by removing the leftmost column with hight $n_1$ as in the right of Fig.~\ref{Young}.
That is, if $Y=[n_1,n_2,\cdots,n_p]$, then $Y'=[n_2, \cdots,n_p]$.
The number of boxes in $Y'$ is $N{-}n_1=K$. This is understood as follows. By removing the leftmost column, 
each value of $p_k=1,2,\cdots, K$ appears precisely once as is clear in table~\ref{Young}.
Note that $p_k$ gives the upper bound of $\ell$ for each fixed $k$.
Now we reinterpret these numbers inside the boxes of $Y'$ as the values of $\ell$. 
Then, for each fixed $\ell$, the value of $k$ is bounded as $k \geq \ell+a'$, where $a'$ is the height of the $\ell$-th box in $Y'$. 
Then we get $p'_\ell=2\ell-q_\ell=\ell-a'$. This is exactly the rule which associates the degrees of poles to the Young diagram $Y'$.

Now that we know $K=p_N$ and the puncture $Y'$, it is easy to reconstruct the linear quiver, using the standard class S technology.  When the original puncture $Y$ is the full puncture, we find the linear quiver \begin{equation}
[\SU(N)_A]-\SU(N{-}1)-\SU(N{-}2)-\cdots - \SU(2)-\SU(1).\label{eq:foo}
\end{equation}  
In the more general case, we can check that it indeed reproduces the quiver given in \eqref{eq:generalquiver}.

\paragraph{Examples.}
Let us consider specific examples where the puncture $Y$ is given by the partition of $N$ as $Y=[N{-}K,1^{K}]$. 
($K=N{-}1$ corresponds to the full puncture.) Then, by the rule discussed above, we have $Y'=[1^K]$.
%\beq
%p_k= \left\{
%\begin{array}{ll}
%k-1, & k \leq K+1 \\
%K, &  k \geq K+1
%\end{array}
%\right.
%\eeq
%and hence
%\beq
%q_\ell=\ell+1,~~p'_\ell=\ell-1,~~~(\ell \leq K).
%\eeq
So, the puncture $Y'$ is a full puncture of the $A_{K-1}$ theory.
In this case \eqref{eq:dualYcurve} represents the curve of the theory which has $N$ simple punctures and one full puncture.
In one dual frame, this theory is realized by the linear quiver
\beq
[\SU(K)] -\SU(K)-\cdots \SU(K)-\SU(K-1)-\cdots-\SU(2)-\SU(1).
\eeq
where the number of $\SU(K)$ is $N{-}K$, and the rightmost $\SU(K)$ has an additional fundamental hypermultiplet to make the theory conformal.
This quiver can also be derived from the general form given above: $w_1=K$, $w_{N{-}K}=1$ and other $w_i$ are zero. This datum determines $v_i$. 

The flavor symmetry of the full puncture is $\SU(K)$ when $K<N{-}1$.  When $K=N{-}1$, the bifundamentals of $[\SU(N{-}1)] -\SU(N{-}1)$
and the additional fundamental of $\SU(N{-}1)$ are combined and the symmetry is enhanced to $\SU(N)$.

\subsection{Higgs branches}\label{sec:higgs}
As a final check, we directly show that the Higgs branch of the linear quiver 
\eqref{eq:foo},
\begin{equation}
[\SU(N)_A]-\SU(N{-}1)-\SU(N{-}2)-\cdots-\SU(2)-\SU(1),
\end{equation} equals that of the $T_N$ theory under the generic mass deformations $\mm_{B,C}$ to $\SU(N)_{B,C}$. 

Let us first study the Higgs branch of the linear quiver.  Note that when $\mm_{B,C}$ are generic, all the bifundamental fields have masses associated to the $\U(1)$ baryon symmetries. 
The only gauge-invariant field uncharged under baryonic symmetries are $\mu_{A}$ that is  the quadratic combination of the leftmost bifundamental, transforming as an adjoint of $\SU(N)_A$. 

Let us give vevs to the adjoint scalars of the vector multiplets of $\SU(k)~(k=2,3,\cdots,N{-}1)$ so that the maximal number of bifundamentals becomes
massless. By appropriate vevs which break the gauge groups as $\SU(k) \to \U(k-1)$, the theory can be reduced to a quiver
\beq
[\SU(N)_A]-\U(N{-}2)-\U(N{-}3)-\cdots-\U(1),
\eeq
where all the bifundamentals are massless.
It is standard that $\tr \mu_A^k=0$ and $\mu^{N{-}1}_A=0$ follow from the F-term conditions, 
see e.g.~Sec.~3.3 of \cite{Gaiotto:2008ak}. 
Thus $\mu_A$ is a nilpotent $N\times N$ matrix which, by complexified $[\SU(N)_A]$ transformations, is conjugate to a block diagonal matrix 
$J_{N{-}1} \oplus J_1$.

Next we study the Higgs branch of the $T_N$ theory under general mass deformations $\mm_{B,C}$. 
We use two different methods.
\paragraph{Higgs branch from chiral rings.}
As already discussed in Sec.~\ref{sec:UV}, the mass deformations kill all operators charged under the Cartan of $\SU(N)_{B,C}$.   In particular, $Q^{i_Ai_Bi_C}$ and its cousins with more indices are all set to zero. The adjoint operators $\mu_{B,C}$ are required to be diagonal, and we do not see conditions on $\mu_A$ yet. 

There are the well-known chiral ring relations \begin{equation}
\tr \mu_A^k = \tr \mu_B^k = \tr \mu_C^k \label{eq:trk0}
\end{equation} for all $k$. Therefore, if we can show $\mu_{B,C}=0$ we get $\tr \mu_A^k=0$.
We already know that $\mu_{B,C}$ are diagonal. 
Then the relations  \eqref{eq:trk0} mean that  we can assume $\mu_B=\mu_C$.
Let us denote their $N$ eigenvalues to be $\mu_{1,\ldots,N}$.
With this, we have the chiral ring relation \begin{multline}
Q^{[i_{A,1}\cdots i_{A,k}][i_{B,1}\cdots i_{B,k}][i_{C,1}\cdots i_{C,k}]}
Q_{[i_{A,1}\cdots i_{A,k}][j_{B,1}\cdots j_{B,k}][j_{C,1}\cdots j_{C,k}]}
=\\
\delta^{[i_{B,1}\cdots i_{B,k}]}_{j_{B,1}\cdots j_{B,k}}
\delta^{[i_{C,1}\cdots i_{C,k}]}_{j_{C,1}\cdots j_{C,k}}
\prod_{i\in \{i_{B,1}\cdots i_{B,k}\}}
\prod_{j\not\in \{i_{C,1}\cdots i_{C,k}\}} 
(\mu_i-\mu_j) 
\end{multline} that follows from \eqref{eq:qrelation3} in Appendix~\ref{sec:higgsrelations}.

We already know that all $Q$ operators are zero. Since $k$ is arbitrary in the relations above, we see that all $\mu_i=0$, forcing $\mu_{B,C}=0$. Therefore $\tr\mu_A^k=0$. Furthermore, from the relation \eqref{eq:qqmm} applied to $A$ and $B$ and using $\mu_B=0$, we get $\mu_A^{N{-}1}=0$.
These are what we wanted.

\paragraph{Higgs branch from SW curve.}
We can also obtain the same result from the Seiberg-Witten curve of the $T_N$ theory \eqref{eq:Ycurve} 
by using the method in \cite{Chacaltana:2012zy,Xie:2014pua}.
Let us make the curve the least singular at the puncture $z=0$. From \eqref{eq:limsing},
this is achieved when all the Coulomb moduli are tuned to be zero, and we get $\phi_k \sim z^{-1}$.
This singularity is the one allowed by a simple puncture corresponding to the partition $[N{-}1,1]$.
Then, we can go to the Higgs branch where the puncture is Higgsed by a nilpotent vev conjugate to $J_{N{-}1} \oplus J_1$.
This branch is exactly given by $\tr \mu_A^k=0$ and $\mu_A^{N{-}1}=0$.

\subsection{From star shaped quiver to linear quiver in 3d}\label{sec:3d}
In three dimensions, the $T_N$ theory has a Lagrangian description in terms of a star-shaped quiver
if we take mirror symmetry~\cite{Benini:2010uu}.
First, we define a superconformal theory $T[\SU(N)]$ as the low energy limit of the quiver~\cite{Gaiotto:2008ak}
\beq
[\SU(N)]-\U(N-1)-\U(N-2)-\cdots-\U(1). \label{eq:TSUNquiver}
\eeq
In addition to the visible flavor $\SU(N)$ symmetry of the Higgs branch, the $T[\SU(N)]$ has another flavor $\SU(N)$ symmetry associated to the Coulomb branch.
The Cartan subalgebra of this Coulomb branch $\SU(N)$ symmetry is generated by the topological currents 
$j_k=\tr F_k$ associated to $\U(k)~(k=1,2,\cdots, N-1)$ gauge groups, where $F_k$ is the field strength of $\U(k)$.
We also define the moment map of the Higgs branch $\SU(N)$ as $M=A_{N-1} \tilde{A}_{N-1}-\frac{1}{N}\tr A_{N-1} \tilde{A}_{N-1}$, 
where $A_k, \tilde{A}_k~(k=1,\cdots,N-1)$ 
are the bifundamentals of $[\SU(N)]-\U(N-1)$ (for $k=N-1$) and $\U(k+1)-\U(k)$.

Now, we take three copies of $T[\SU(N)]$, which we denote as $T[\SU(N)]_A$, $T[\SU(N)]_B$ and $T[\SU(N)]_C$ respectively.
Then the mirror of the 3d $T_N$ theory is obtained by coupling the Higgs branch $\SU(N)$ symmetries of $T[\SU(N)]_{A,B,C}$
 to a single gauge group $\SU(N)$. This gives the star-shaped quiver.
 The Coulomb branch $\SU(N)$ symmetries of $T[\SU(N)]_{A,B,C}$ become the mirror of the flavor symmetries of the $T_N$ theory.
 
 When we add mass terms to the Cartan of $\SU(N)_{X} ~(X=A,B,C)$ of the $T_N$ theory, 
 they become FI parameters of the gauge groups $\U(k)~(k=1,\cdots,N-1)$ of the corresponding $T[\SU(N)]_X$ in the mirror side.
If these FI parameters are generic, the moment map $M_X$ gets a diagonal vev (see Sec.~3.3 of \cite{Gaiotto:2008ak}).
This vev breaks the $\SU(N)$ gauge group at the center of the star-shaped quiver to $\U(1)^{N-1}$.
Therefore, by generic deformation of $\SU(N)_B \times \SU(N)_C$ and integrating out massive degrees of freedom,
we get a system in which the $T[\SU(N)]_A$ survives and the Cartan of its Higgs branch $\SU(N)$ is gauged by the $\U(1)^{N-1}$.

By taking the mirror of the above system again, we get the low energy theory of the mass-deformed $T_N$ in the original description.
The $T[\SU(N)]_A$ is self-dual under the mirror symmetry, but its Higgs and Coulomb branches are exchanged.
The gauging of the Higgs branch symmetry by $\U(1)^{N-1}$ becomes the gauging of the Coulomb branch symmetry by the mirror symmetry.
As mentioned above, this Coulomb branch symmetry is generated by the topological currents $j_k=\tr F_k$, so gauging this symmetry
gives Chern-Simons couplings $A'_k \wedge \tr F_k$, where $A'_k$ are the gauge fields of $\U(1)^{N-1}$.
These Chern-Simons couplings make all the $\U(1)$ gauge fields massive, including the $\U(1)_k \subset \U(k)$ subgroups.
Therefore, we finally get a quiver in which all the gauge groups in \eqref{eq:TSUNquiver} become special unitary $\SU$ groups instead of unitary 
$\U$ groups. This is exactly what we wanted. 

The mass terms of the bifundamentals are generated by integrating out the massive fields in the 
superpotential  which is schematically given by
\beq
W \sim  \sum_k \Phi'_k  \tr \Phi_k +\sum_k \tr [\Phi_k(A_k\tilde{A}_k-A_{k-1}\tilde{A}_{k-1})] + \tr\Phi'(\vev{M}_B+\vev{M}_C),
\eeq
where $\Phi'=(\Phi'_k)_{1 \leq k \leq N-1}$ and $\Phi_k$ are the adjoint chiral fields of $\U(1)^{N-1}$ and $\U(k)$, respectively. 
The first term in the above superpotential is the supersymmetric partner of the Chern-Simons terms, while the second and third terms are the usual 
couplings of the adjoint chiral fields to hypermultiplets.

It is easy to generalize this analysis to the 3d theory corresponding to the three-punctured sphere with two full punctures and one puncture of type $Y$. The 3d mirror to this theory is obtained by taking two copies of $T[\SU(N)]$ theory and one theory $T^Y[\SU(N)]$ as introduced in \cite{Gaiotto:2008ak}, and gauging the common $\SU(N)$ flavor symmetry. Now we give FI terms to the $\SU(N)^2$ symmetry of the two copies of $T[\SU(N)]$ theory. Proceeding as before, we end up with $\U(1)^{N-1}$ gauge fields gauging the Cartan of the $\SU(N)$ symmetry of $T^Y[\SU(N)]$. Now we perform the 3d mirror again. The mirror of $T^Y[\SU(N)]$ is the 3d quiver of the form 
\eqref{eq:generalQ}, but with $\U(v_i)$ gauge groups instead of $\SU(v_i)$ gauge groups.  The $\U(1)^{N-1}$ gauge fields now couple to the topological charge of the $\U(1)$ parts of $\U(v_i)$ gauge groups, effectively eliminating them.  We thus end up exactly with the quiver of the form \eqref{eq:generalQ} with $\SU(v_i)$ gauge symmetries. 

\section*{Acknowledgements}
HH and YT would like to thank Institute for Advanced Study for hospitality during Prospects in Theoretical Physics 2014, where this project was initiated. HH would also like to thank Mainz Institute for Theoretical Physics for hospitality and its partial support during a part of this work. The work of HH is supported by the REA grant agreement PCIG10-GA-2011-304023 from the People Programme of FP7 (Marie Curie Action), the grant FPA2012-32828 from the MINECO, the ERC Advanced Grant SPLE under contract ERC-2012-ADG-20120216-320421 and the grant SEV-2012-0249 of the ``Centro de Excelencia Severo Ochoa'' Programme.
The work of YT is  supported in part by JSPS Grant-in-Aid for Scientific Research No. 25870159,
and in part by WPI Initiative, MEXT, Japan at IPMU, the University of Tokyo.
The work of KY is supported in part by NSF Grant PHY-0969448.

\appendix
\section{Higgs branch chiral ring relations of the $T_N$ theory}\label{sec:higgsrelations}
The Higgs branch operators of the $T_N$ theory are generated by 
\beq
&(\mu_A)^{i_A}_{j_A}, ~(\mu_B)^{i_B}_{j_B},~(\mu_C)^{i_C}_{j_C}, \nonumber \\
&Q^{[i_{A,1},\cdots, i_{A,k}],[i_{B,1},\cdots, i_{B,k}],[i_{C,1},\cdots, i_{C,k}]}~(k=1,\cdots,N{-}1)
\eeq
where $i_A$, $i_B$, $i_C$ etc. are indices of $\SU(N)_A$, $\SU(N)_B$ and $\SU(N)_C$ respectively, 
and $[i_1 , \cdots, i_k]$ means that the indices are anti-symmetrized. 
The $\mu_{A,B,C}$ are in the adjoint representations of $\SU(N)_{A,B,C}$, and $Q$ are in the representation
$(\wedge^k, \wedge^k, \wedge^k)$ of $\SU(N)_A \times \SU(N)_B \times \SU(N)_C$, where $\wedge^k$
means the $k$-th anti-symmetric representation of $\SU(N)$.  The relation between $Q^{[i_{A,1}\cdots j_{A,k}]  [i_{B,1}\cdots i_{B,k}] [i_{C,1}\cdots i_{C,k}]  } $
and $Q_{[i_{A,1}\cdots j_{A,k}]  [i_{B,1}\cdots i_{B,k}] [i_{C,1}\cdots i_{C,k}]  } $ is
\begin{multline}
Q_{[i_{A,1}\cdots j_{A,k}]  [i_{B,1}\cdots i_{B,k}] [i_{C,1}\cdots i_{C,k}]  } \\
=\frac{1}{(N{-}k)!^3} \epsilon_{i_{A,1} \cdots i_{A,N}}\epsilon_{i_{B,1} \cdots i_{B,N}} \epsilon_{i_{C,1} \cdots i_{C,N}}
Q^{[i_{A,k+1}\cdots i_{A,N}]  [i_{B,k+1}\cdots i_{B,N}] [i_{C,k+1}\cdots i_{C,N}]  }.
\end{multline}

We have \begin{equation}
\tr \mu_A^k = \tr \mu_B^k = \tr \mu_C^k \label{eq:trk}
\end{equation} for all $k$. Let us define $v_k$ via
\begin{equation}
P(x)=\det(x-\mu_X)=\sum_{k=0}^N v_k x^{N{-}k}. 
\end{equation} where $X$ can be either $A$, $B$ or $C$. 

Then the following relation was described in \cite{Maruyoshi:2013hja}:
\begin{equation}
Q^{i_A i_B i_C} Q_{i_A j_B j_C}  = \sum_{l=0}^N  v_l \sum_{m=0}^{N{-}l-1} (\mu_B^{N{-}l-1-m})^{i_B}_{j_B} (\mu_C^m)^{i_C}_{j_C}. \label{eq:qqmm}
\end{equation} Here we contracted the index $i_A$; we of course have the corresponding identities when the indices $i_B$ or $i_C$ are contracted.

Suppose that the vevs of $\mu_{A,B,C}$ are given as 
\beq
\mu_A=\mu_B=\mu_C=\diag(\mu_1,\cdots,\mu_N), \label{eq:diagmu}
\eeq
where $\mu_1,\cdots, \mu_N$ are generic. 
The complex dimension of the subspace of the Higgs branch under this condition on $\mu_{A,B,C}$ is $N{-}1$. 

It is strongly believed that when $\mu_{A,B,C}$ are generic and diagonal as above,  the only nonzero components of 
$Q^{[i_{A,1}\cdots i_{A,k}]  [i_{B,1}\cdots i_{B,k}] [i_{C,1}\cdots i_{C,k}]  }$ and $Q_{[i_{A,1}\cdots i_{A,k}]  [i_{B,1}\cdots i_{B,k}] [i_{C,1}\cdots i_{C,k}]  }$ are  
\beq
Q^{[i_{1}\cdots i_{k}]  [i_{1}\cdots i_{k}] [i_{1}\cdots i_{k}]  } =q^{[i_{1}\cdots i_{k}]},~~~
Q_{[i_{1}\cdots i_{k}]  [i_{1}\cdots i_{k}] [i_{1}\cdots i_{k}]  } =q_{[i_{1}\cdots i_{k}]}.\label{eq:diagQ}
\eeq

From section 2 of \cite{Maruyoshi:2013hja}, we have
\beq
q^i q_i=\prod_{j \neq i} (\mu_i-\mu_j).\label{eq:qrelation2}
\eeq This already provides $N$ dimensions with $\mu_{A,B,C}$ fixed to be diagonal. 
The complex dimension of the Higgs branch of the $T_N$ theory is given by $2(N-1)+3N(N-1)$, where $3N(N-1)$ comes from actions of
complexified $\SU(N)_{A,B,C}$ to the above diagonal $\mu_{A,B,C}$.
To reproduce the correct dimensions, $q^{[i_{1}\cdots i_{k}]}$ need to be given by $q^i$ and $\mu_i$, and 
there must be one relation among $q^i$. 
A sensible guess is then 
\beq
&q^{i_1}\cdots q^{i_k} =  q^{[i_{1}\cdots i_{k}]  } \prod_{1\leq a<b \leq k} (\mu_{i_a} -\mu_{i_b}), \nonumber \\
&q_{i_1}\cdots q_{i_k} =  q_{[i_{1}\cdots i_{k}]  } \prod_{1\leq a<b \leq k} (\mu_{i_a} -\mu_{i_b}). \label{eq:qrelation1}
\eeq
This equation was obtained for $k=N-1$ and $k=N$ in \cite{Maruyoshi:2013hja} with $q^{[i_{1}\cdots i_{N}] }$ interpreted to be constant, 
and hence this is a natural generalization for arbitrary $k$. 
Combining \eqref{eq:qrelation1} and \eqref{eq:qrelation2}, and assuming $\mu_1, \cdots, \mu_N$ are generic, we get
\beq
q^{[i_{1}\cdots i_{k}]  }q_{[i_{1}\cdots i_{k}]  }={(-1)^{\frac{1}{2}k(k-1)}} \prod_{i \in I, ~j \not\in I} (\mu_i -\mu_j),\label{eq:qrelation3}
\eeq
where $I=\{ i_1,\cdots, i_k\}$.

A candidate chiral ring relation that reduces to \eqref{eq:qrelation3} when $\mu_{A,B,C}$ are generic can be written down as follows: the left hand side is, in general, given by \begin{multline}
L_{[i_{B,1}\cdots i_{B,k}][j_{B,1}\cdots j_{B,N{-}k}]
[i_{C,1}\cdots i_{C,k}][j_{C,1}\cdots j_{C,N{-}k}]}\\
:=Q_{[i_{A,1}\cdots i_{A,k}][i_{B,1}\cdots i_{B,k}][i_{C,1}\cdots i_{C,k}]}
Q_{[j_{A,1}\cdots j_{A,N{-}k}][j_{B,1}\cdots j_{B,N{-}k}][j_{C,1}\cdots j_{C,N{-}k}]}
\epsilon^{i_{A,1}\cdots i_{A,k}j_{A,1}\cdots j_{A,N{-}k}}.
\end{multline}

A combination of $\mu_B$ and $\mu_C$ with the correct index structure, the scaling dimension, that reduces to the right hand side of \eqref{eq:qrelation3} is then \begin{equation}
\left[\prod_{i\in \{1,\ldots,k\}}
\prod_{j\in \{1,\ldots,N{-}k\}}  
(\mu_{B,i} - \mu_{C,j+k} )\right]
\epsilon_B \epsilon_C . \label{eq:CRguess}
\end{equation}
Here, $\epsilon_{B,C}$ is the epsilon symbol for $\SU(N)_{B,C}$ regarded as the standard element of $\wedge^N V_{B,C} \subset \otimes^N V_{B,C}$ where $V_{B,C}$ are the $N$ dimensional spaces on which $\SU(N)_{B,C}$ act, and $\mu_{X,i}$ is the $\mu_X$ regarded as acting on $i$-th factor of $\otimes^N V_X$. The total anti-symmetry of $\epsilon_{B,C}$ is partially broken by the actions of $(\mu_B-\mu_C)$'s, and \eqref{eq:CRguess} takes values in $\wedge^{k} V_B \otimes \wedge^{N{-}k}V_B \otimes \wedge^{k} V_C \otimes \wedge^{N{-}k}V_C  $.

When $\mu$'s are given as \eqref{eq:diagmu}, one can see that the components of \eqref{eq:CRguess} are given as
\beq
&L_{[i_{B,1}\cdots i_{B,k}][j_{B,1}\cdots j_{B,N{-}k}]
[i_{C,1}\cdots i_{C,k}][j_{C,1}\cdots j_{C,N{-}k}]} \nonumber \\
=&\epsilon_{i_{B,1}\cdots i_{B,k}j_{B,1}\cdots j_{B,N{-}k}}
\epsilon_{i_{C,1}\cdots i_{C,k}j_{C,1}\cdots j_{C,N{-}k}} 
\prod_{i\in \{i_{B,1}\cdots i_{B,k}\}}
\prod_{j\in \{j_{C,1}\cdots j_{C,N{-}k}\}} 
(\mu_i-\mu_j) .
\eeq
This has the desired properties that; (i) it is nonzero if and only if $\{ j_{B,1}\cdots j_{B,N{-}k} \}$ is the complement of $\{i_{B,1}\cdots i_{B,k}\}$
(and similarly for $C$) 
and $\{i_{B,1}\cdots i_{B,k}\}=\{i_{C,1}\cdots i_{C,k}\}$ as required by \eqref{eq:diagQ}, 
and (ii) it has the correct anti-symmetric properties of the indices.

%Hopefully this reduces to \eqref{eq:qqmm} when $k=1$ ... 

\bibliographystyle{ytphys}
\small\baselineskip=.9\baselineskip
\let\bbb\bibitem\def\bibitem{\itemsep1pt\bbb}
\bibliography{ref}
\end{document}